\documentclass[preprintnumbers, superscriptaddress, showpacs, nofootinbib, amsfonts, amsmath, amssymb, aps, prd, notitlepage, floatfix]{revtex4-2}

\usepackage[utf8]{inputenc}
\usepackage{amsmath,amssymb,amsfonts,mathrsfs,bbold}
\usepackage{yfonts}
\usepackage{graphicx}
\usepackage[colorlinks=true,citecolor=blue,linkcolor=blue,urlcolor=blue]{hyperref}
\usepackage[dvipsnames]{xcolor}
\usepackage{lipsum}
\usepackage{slashed}
\usepackage[normalem]{ulem}
\usepackage{url}
\usepackage{physics}

\usepackage{amssymb, bm, amsmath, graphicx}
\usepackage{enumerate}
\usepackage{amsfonts}
\usepackage{mathtools}
\usepackage{latexsym}
\usepackage{epstopdf}
\usepackage{subfigure}
\usepackage{tikz}
\usetikzlibrary{decorations.pathmorphing, decorations.markings, shapes.geometric, shapes.misc, calc, math}
\tikzstyle{gluon}=[decorate, decoration={coil,aspect=0.8, amplitude=1.5pt,  segment length=3pt}]
\usepackage{stackrel}
\usepackage[most]{tcolorbox}
\usepackage{dsfont}

\def\sec#1{{Sec.~\ref{#1}}}
\def\eq#1{{Eq.~(\ref{#1})}}
\def\fig#1{{Fig.~\ref{#1}}}
\newcommand{\ben}{\begin{eqnarray*}}
\newcommand{\een}{\end{eqnarray*}}

\newcommand{\as}{\alpha_s}

\newcommand{\pd}{\partial}

\allowdisplaybreaks

\begin{document}

\title{Novel Cross Section Ratios as Possible Signals of Saturation in UPCs}


    \makeatletter  
\def\@fnsymbol#1{\ensuremath{\ifcase#1\or *\or \dagger\or \ddagger\or
   \mathsection\or \mathparagraph\or \|\or **\or \dagger\dagger
   \or \ddagger\ddagger \or \mathsection\mathsection \else\@ctrerr\fi}}
    \makeatother

\author{Yuri~V.~Kovchegov}
\email{kovchegov.1@osu.edu}
\affiliation{Department of Physics, The Ohio State University, Columbus, Ohio 43210, USA}

\author{Huachen Sun}
\email{sun.2885@buckeyemail.osu.edu}
\affiliation{Department of Physics, The Ohio State University, Columbus, Ohio 43210, USA}

\author{Zhoudunming Tu}
\email{zhoudunming@bnl.gov}
\affiliation{Department of Physics, Brookhaven National Laboratory, Upton, New York 11973, USA}


\begin{abstract}
We propose new cross section ratios  
in ultra-peripheral A$+$A and $p+$A collisions (UPCs) as a potential experimental signal of saturation physics. We consider the ratio $R_1$ of elastic vector meson photoproduction cross section to the inclusive hadron or jet photoproduction cross section. The ratio can be measured in the $\gamma +$A and $\gamma + p$ collisions taking place in the UPCs. We label the ratios $R_1 ({\rm A})$ and $R_1 (p)$, respectively. Constructing the double ratio $R_{\rm{UPC}} ({\rm A}) = R_1 ({\rm A})/R_1 (p)$, and performing a small-$x$ calculation both in the quasi-classical approximation and by including small-$x$ evolution, we observe that $R_{\rm{UPC}} ({\rm A})$ exhibits a markedly different dependence on the nuclear atomic number inside and outside the saturation region. This result indicates that $R_{\rm{UPC}} ({\rm A})$ and  $R_1 ({\rm A})$ measurements in UPCs may help in the experimental searches for the evidence of saturation physics. 
\end{abstract}

\date{\today}
\maketitle
\tableofcontents

%
\section{Introduction}
%

\subsection{Saturation physics}


Understanding Quantum Chromodynamics (QCD) in high-energy scattering is one of the open problems in physics. An interesting phenomenon one expects to find at very high energy is the saturation of gluon and quark densities in the small Bjorken $x$ region of the hadronic and nuclear wave functions \cite{Gribov:1984tu}. (Small values of Bjorken $x$ are probed in high-energy collisions.) The quark and gluon parton distribution functions (PDFs) are known to grow with decreasing $x$. However, this rise of PDFs cannot continue indefinitely: larger number of partons leads to higher parton densities in the hadronic or nuclear wave functions, which, in turn, lead to parton mergers becoming important, slowing down the growth of the PDF with decreasing $x$. The regime where the parton mergers compensate for the parton splittings is known as {\it the parton saturation} (see Refs.~\cite{Gribov:1984tu, Iancu:2003xm, Weigert:2005us, JalilianMarian:2005jf, Gelis:2010nm, Albacete:2014fwa, Kovchegov:2012mbw, Morreale:2021pnn} for reviews). Unambiguously  identifying  saturation signatures in the experimental data is one of the main challenges of the modern hadronic physics, comprising a large part of the experimental program for the future Electron-Ion Collider (EIC) \cite{Accardi:2012qut,Boer:2011fh,Proceedings:2020eah,AbdulKhalek:2021gbh}.

Theoretical description of parton saturation usually involves the linear Balitsky--Fadin--Kuraev--Lipatov (BFKL) \cite{Kuraev:1977fs,Balitsky:1978ic} and non-linear Balitsky--Kovchegov (BK) \cite{Balitsky:1995ub,Balitsky:1998ya,Kovchegov:1999yj,Kovchegov:1999ua} and Jalilian-Marian--Iancu--McLerran--Weigert--Leonidov--Kovner (JIMWLK) \cite{Jalilian-Marian:1997dw,Jalilian-Marian:1997gr,Weigert:2000gi,Iancu:2001ad,Iancu:2000hn,Ferreiro:2001qy} evolution equations in $x$. While the linear BFKL evolution predicts the rise of PDFs at small $x$, it does not contain saturation effects, and leads to cross sections growing as a power of the center-of-mass energy, violating unitarity at very high energies. Unitarity (in the sense of adhering to the black disk limit) is restored when non-linear (parton merger) effects are taken into account: this is accomplished by using the BK/JIMWLK evolution equations. The initial conditions for the non-linear evolution equations are given by the quasi-classical Glauber--Gribov--Mueller (GGM)/McLerran--Venugopalan (MV) model \cite{Mueller:1989st,McLerran:1993ni,McLerran:1993ka,McLerran:1994vd}, which includes all-orders of multiple scatterings of the projectile in the target. The above description is valid for the scattering of a compact projectile on a large target, as seen in deep inelastic scattering (DIS) experiments on nuclei (and, possibly, on protons) and proton--nucleus ($p+$A) collisions at high energies. The saturation formalism is also referred to as the color glass condensate (CGC). 

The non-linear small-$x$ evolution combined with the GGM-MV initial conditions dynamically generates a momentum scale intrinsic to the hadronic and nuclear wave functions at small $x$ -- {\it the saturation scale} $Q_s$. This scale designates the transition from the low-density regime (with momentum scales $Q > Q_s$) to the saturated high-density regime (for $Q \lesssim Q_s$). The saturation scale grows with the atomic number $A$ of the nucleus and with decreasing $x$, approximately as $Q_s^2 \propto A^{1/3} \, (1/x)^\lambda$ with the power $\lambda \approx 0.3$. This means that, in high-energy collisions (corresponding to low values of $x$) and/or for collisions involving sufficiently large nuclei (with large $A$), the saturation scale should get large. When it gets sufficiently above the QCD confinement scale, $Q_s \gg \Lambda_\textrm{QCD}$, the corresponding strong coupling constant becomes small, $\as (Q_s^2) \ll 1$, allowing for a weak-coupling description of the small-$x$ hadronic and nuclear wave functions and, therefore, justifying the application of perturbative techniques to the problem. 

Tantalizing evidence for observation of the saturation regime has been collected by accelerators around the world (see e.g. \cite{Accardi:2012qut} for a summary of the experimental results pertaining to saturation physics, along with the more recent data \cite{CMS:2022nnw,ALICE:2023gcs,ALICE:2023jgu,STAR:2021fgw}). It is widely anticipated that the future EIC, due to its reach in $x$ and $Q^2$, its unprecedented high luminosity, and a variety of nuclear targets, would be able to seal the discovery of the saturation phenomenon. However, with the EIC completion being projected for the early 2030s, in the meantime, the hadronic physics community must make every effort to study for potential saturation signals at the existing high-energy accelerators, such as the Relativistic Heavy-Ion Collider (RHIC) and the Large Hadron Collider (LHC). In this respect, a particularly promising channel appears to be the ultra-peripheral collisions (UPCs) \cite{Bertulani:2005ru,Baltz:2007kq,Klein:2019qfb,Klein:2020fmr,Arslandok:2023utm}. As we describe below, in the UPCs one usually collides two large nuclei (A$+$A) or a proton and a nucleus ($p+$A) at very large impact parameters, much larger than the sum of the radii for the scattering particles. The interaction happens due to an almost-real photon coming from one of the nuclei in A$+$A or from the nucleus in $p+$A collisions (the projectile)  interacting with the other nucleus or the proton (the target) by fluctuating into a quark-antiquark ($q \bar q$) dipole, which then scatters on the target predominantly via $t$-channel gluon exchanges (at high energy). Thus, the UPCs provide one an opportunity to perform a small-$x$ DIS-type of experiment, either on the proton (in $p+$A) or nuclear (in A$+$A) targets, except with an almost real photon ($Q^2 \approx 0$). The UPCs thus allow us to study the scattering amplitude of a $q \bar q$ dipole on a target, searching for the signals of saturation. The $q \bar q$ dipole scattering amplitude is an essential component of saturation physics, directly entering the BK evolution equation. The aim of this work is to propose a new saturation signal that can be studied in the UPCs at RHIC and LHC.

\subsection{Double ratio in DIS}

One of the proposed key saturation signals at the  
EIC is  
the fraction of diffractive DIS with respect to the total DIS cross section in electron-nucleus ($e+$A) collisions~\cite{Accardi:2012qut}. As this ratio is found to be $\sim13\%$ at HERA in electron-proton ($e+p$) collisions~\cite{Abramowicz:1998ii}, saturation calculations predict that this ratio will be enhanced (larger than 13\%) in $e+$A collisions. To quantify this enhancement, a double ratio observable was introduced between $e+$A and $e+p$ collisions and saturation models predict it to be above one. The general idea behind this observable is that in the black disk limit the elastic cross section comprises $50\%$ of the total cross section. Indeed, for a $q \bar q$ dipole scattering on a target the total and elastic scattering cross sections can be written as 
\begin{subequations}
    \begin{align}
        & \sigma_{\rm tot} = 2 \, \int d^2 b_\perp \, N ({\mathbf r}, {\mathbf b}, Y), \\ 
        & \sigma_{\rm el} = \int d^2 b_\perp \, \left[ N ({\mathbf r}, {\mathbf b}, Y) \right]^2 ,
    \end{align}
\end{subequations}
where $N ({\mathbf r}, {\mathbf b}, Y)$ is the (imaginary part of the) forward scattering amplitude for the dipole of transverse separation $\mathbf r$ interacting with the target at the impact parameter $\mathbf b$ with the dependence on the center-of-mass energy squared $s$ entering through the rapidity variable $Y = \ln (s \, r_\perp^2) \approx \ln (1/x)$ \cite{Kovchegov:1999yj}. (In our notation, two-dimensional transverse vectors are labeled $\mathbf{v} = (v^x, v^y)$, their magnitude is $v_\perp = |{\mathbf v}|$, while the two-dimensional integration measure is $d^2 v_\perp$. For the transverse momenta $p_T = |\mathbf{p}|$ and the integration measure is $d^2 p_T$.) In the black disk limit $N=1$, such that $\sigma_{\rm el} / \sigma_{\rm tot}  = 1/2$ (see e.g. \cite{Kovchegov:2012mbw} for details). Away from the black disk limit, where $N \ll 1$, the elastic to total cross sections ratio is small, $\sigma_{\rm el} / \sigma_{\rm tot} \ll 1$. Saturation physics is responsible for the approach to and the onset of the black disk limit, leading to the anticipated increase of diffractive (elastic) cross section $\sigma_{\rm diff}$ as a fraction of the total cross section $\sigma_{\rm tot}$,  both with increasing nuclear size ($A$) and with decreasing $x$. 

The relevant double ratio is \cite{Accardi:2012qut}
\begin{align}
    R_{\rm {EIC}} = \frac{\left[\left(d \sigma_{\rm diff}/d M_X^2 \right) /\sigma_{\rm tot}\right]_{\rm e+A}} {\left[ \left( d \sigma_{\rm diff}/d M_X^2 \right)/\sigma_{\rm tot}\right]_{\rm e+p}},
\end{align}
where $M_X^2$ is the invariant mass of the diffractively produced hadrons. Saturation calculations generally predict $R_{\rm {EIC}} > 1$.

\begin{figure}[thb]
\includegraphics[width=3.0in]{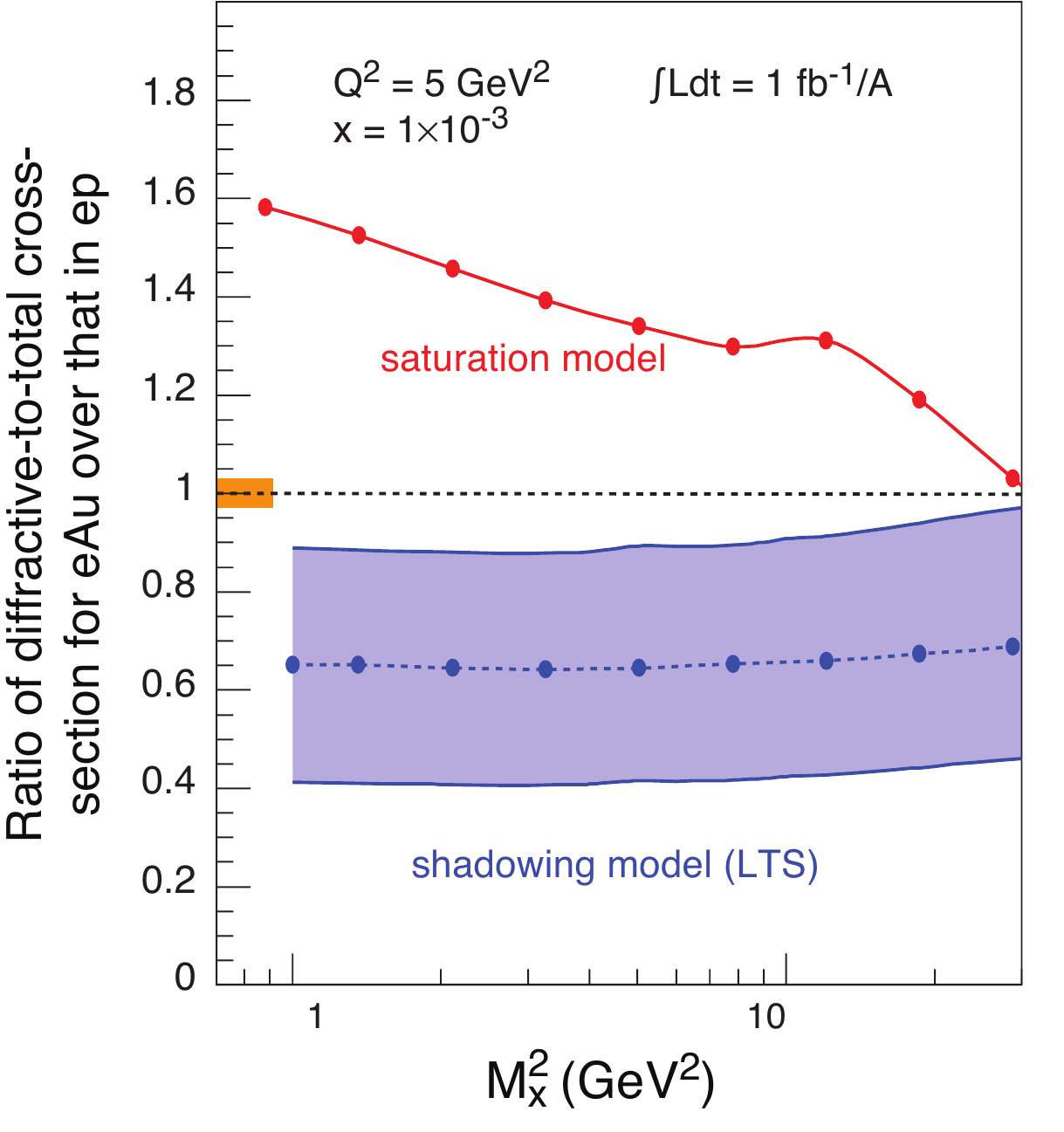}
  \caption{ \label{fig:figure_0} The ratio of diffractive over total cross-section for DIS on gold normalized
to DIS on the proton plotted for different values of $M^{2}_{X}$, the mass squared of hadrons produced in
the collisions, for models assuming saturation and non-saturation. The statistical error bars are
too small to depict and the projected systematic uncertainty for the measurements is shown by
the orange bar. The theoretical uncertainty for the predictions of the LTS model is shown by the grey band. Reprinted from \cite{Accardi:2012qut}.}
\end{figure}

Another reason this double ratio will likely be one of the key
measurements at the EIC is that it may provide maximum distinguishing power between the saturation calculations \cite{Buchmuller:1996xw,Kovchegov:1999kx,Kowalski:2007rw,Kowalski:2008sa,Le:2021qwx,Lappi:2023frf} and 
other theoretical approaches, e.g., the leading-twist nuclear shadowing (LTS) model \cite{Kopeliovich:1999am, Kopeliovich:2002yv, Frankfurt:2003gx, Frankfurt:2011cs}. 
Nuclear shadowing, as considered in \cite{Frankfurt:2003gx,Frankfurt:2011cs}, is expected to cause the fraction of diffractive over total DIS cross sections to decrease in $e+$A with respect to the $e+p$ case, 
leading to a smaller than one double ratio, $R_{\rm {EIC}} < 1$. This is qualitatively different from the prediction of saturation models. The two theoretical predictions for the double ratio to be measured at EIC, coming from saturation physics and the leading-twist shadowing model, are shown in Fig.~\ref{fig:figure_0} \cite{Accardi:2012qut}, where the double ratio $R_{\rm {EIC}}$ is plotted versus the invariant mass of the diffractively produced hadrons $M_X^2$ for fixed values of $x$ and $Q^2$, as indicated in the plot legend.

This is a truly unique measurement that cannot be done at any facility other than the EIC. It requires both $e+p$ and $e+$A DIS measurements at the same center-of-mass energy (or in the same kinematics) and with a large rapidity gap. In addition, most of the systematic uncertainty related to the detector will cancel in this ratio.   

\subsection{A new double ratio in UPCs}
In UPCs, although the same double ratio $R_{\rm {EIC}}$ as proposed above at the EIC cannot be measured, a similar double ratio is possible. First, the ratio between observables related to a proton and a heavy nucleus is possible, because UPC measurements can be done in both $\rm{A+A}$ and $p+\rm{A}$ collisions. In $p+\rm{A}$ collisions, due to the large charge difference between the proton and heavy ion, the dominant process is $\gamma+p$ scattering, where photons are emitted by the heavy ion. Second, the ratio between diffractive DIS and total cross section cannot directly transform into a UPC measurement. Experimentally, rapidity gap events in UPCs are very difficult to measure and the event kinematics cannot be determined due to 
limited acceptance. From a theory perspective, all UPC processes are in the photoproduction limit ($Q^2 \approx 0$) where no hard scale is available for a reliable calculation, unless, at small $x$ and for large nuclei, the saturation scale $Q_s$ becomes large enough. 

While it is hard to replicate the ratio between diffractive DIS and total DIS cross section in UPCs, we propose to replace it with a ratio between the diffractive vector-meson (VM) photoproduction and inclusive particle or jet photoproduction. The idea is similar to the above: as we will show below, the elastic vector meson production cross section is quadratic in the dipole amplitude $N$, while the inclusive particle production is linear in $N$ \cite{Kovchegov:1998bi}, such that their ratio 
should be sensitive to the approach to the black disk limit. 
Here the hard scale of the diffractive probe comes from the mass of the VM, e.g., the $J/\psi$ meson, and from the transverse momentum of the jet or charged particle in inclusive photoproduction, in addition to the saturation scale $Q_s$. The new double ratio is defined as
\begin{align}\label{eq3}
    R_{\rm{UPC}} = \rm{\frac{\left[ \sigma^{VM}_{el}/\left(d\sigma^{jet}_{inclusive}/d^2 p_T \right) \right]_{A+A}} {\left[\sigma^{VM}_{el}/\left( d \sigma^{jet}_{inclusive}/d^2 p_T \right) \right]_{p+A}}}.
\end{align}

Indeed, the ratio $R_{\rm{UPC}}$ is not quite the same as $R_{\rm {EIC}}$. In particular, one may argue that $\sigma_{\rm inclusive} = \langle n \rangle \, \sigma_{\rm inel}$ with $\langle n \rangle$ the average multiplicity of the produced hadrons and $\sigma_{\rm inel}$ the total inelastic cross section. In a (perhaps hypothetical) case when both the nucleus and the proton are scattering near the black disk limit, where $\sigma_{\rm inel} \approx (1/2) \, \sigma_{\rm tot}$ and $R_{\rm {EIC}} \approx 1$, one can approximate $R_{\rm{UPC}} \approx R_{\rm {EIC}} \, \langle n \rangle_{\gamma + {\rm p}} / \langle n \rangle_{\gamma + {\rm A}} \approx \langle n \rangle_{\gamma + {\rm p}} / \langle n \rangle_{\gamma + {\rm A}}$. Since the particle multiplicities are different in $\gamma+p$ and $\gamma+$A collisions, $\langle n \rangle_{\gamma + {\rm p}} / \langle n \rangle_{\gamma + {\rm A}} < 1$, we see that $R_{\rm{UPC}}<1$ in this example. Therefore, we cannot expect that $R_{\rm{UPC}}$ is necessarily always greater than one in the saturation picture, making this ratio different from $R_{\rm {EIC}}$. 

However, what we will show below is that the $A$-scaling of $R_{\rm{UPC}}$ is different inside and outside the saturation region. That is, we will demonstrate that the $A$-dependence of $R_{\rm{UPC}}$ changes depending on whether the produced vector meson is relatively small ($J/\psi$) or large ($\rho$), and also depending on the transverse momentum $p_T$ of the produced hadron or jet in the inclusive cross section being larger than or smaller than the saturation scale $Q_s$. We, therefore, expect that the $A$-dependence of $R_{\rm{UPC}}$ would be of great interest in UPCs, especially since the center-of-mass energy reach can be much higher in the UPCs than that at the EIC. In fact, $R_{\rm{UPC}}$ measurements may lay the groundwork for the future $R_{\rm {EIC}}$ at the EIC, possibly providing necessary complimentary information needed to seal the saturation discovery at the EIC.

\subsection{Outline}

The paper is structured as follows. After giving a brief introduction to UPCs in Sec.~\ref{sec:UPCintro}, we will re-derive the leading-order (LO) expressions for the elastic vector meson production and the total hadron/jet production cross section in the saturation/small-$x$ formalism in Sections~\ref{sec:elastic} and \ref{sec:inclusive}, respectively. While a next-to-leading order (NLO) calculation for elastic vector meson production exists \cite{Mantysaari:2022kdm}, and will need to be utilized in the future more detailed phenomenological studies, our goal is to explore the double ratio $R_{\rm{UPC}}$ at the more qualitative level, trying to identify the difference between its $A$-dependence inside and outside the saturation region. For such a goal, the LO expression for the cross sections should be sufficient. 

In Sec.~\ref{sec:classical} we evaluate the double ratio $R_{\rm{UPC}}$ using the GGM-MV approximation for the dipole amplitude $N$, obtaining different $A$-dependence for $R_{\rm{UPC}}$ inside and outside the saturation region for the $\rho$ and $J/\psi$ meson production (in the numerator of the ratio). We repeat the calculation in Sec.~\ref{sec:evolution} for the dipole amplitude given by the geometric scaling solution of the LO BK equation \cite{Iancu:2002tr,Mueller:2002zm,Gribov:1984tu,Munier:2003sj,Levin:2001cv}, reaching similar conclusions as in the quasi-classical case. Based on our calculations, we discuss the future opportunities to measure $R_{\rm{UPC}}$ in Sec.~\ref{sec:future}, before summarizing our results in Sec.~\ref{sec:conclusions}.

%
\section{Ultra-Peripheral Collisions}
%
\label{sec:UPCintro}

Ultra-peripheral collisions (UPCs) represent a fascinating experimental opportunity at heavy-ion collider facilities, offering a captivating avenue for scientific exploration of the fundamental structure of nucleon and nuclei. In these collisions, heavy ions such as gold (Au) or lead (Pb) approach each other with impact parameters far exceeding the size of the ions themselves, e.g., $b > 2R_{A}$. See Fig.~\ref{fig:figure_1} for an illustration. This results in an electromagnetic interaction dominated by the exchange of virtual photons, where the photon probe interacts with the other nucleus. Therefore, the UPCs are similar to the electron-ion scattering experiments with quasi-real photon beams. Here we list a few of the important features of UPCs:

\begin{figure}[thb]
\includegraphics[width=6.0in]{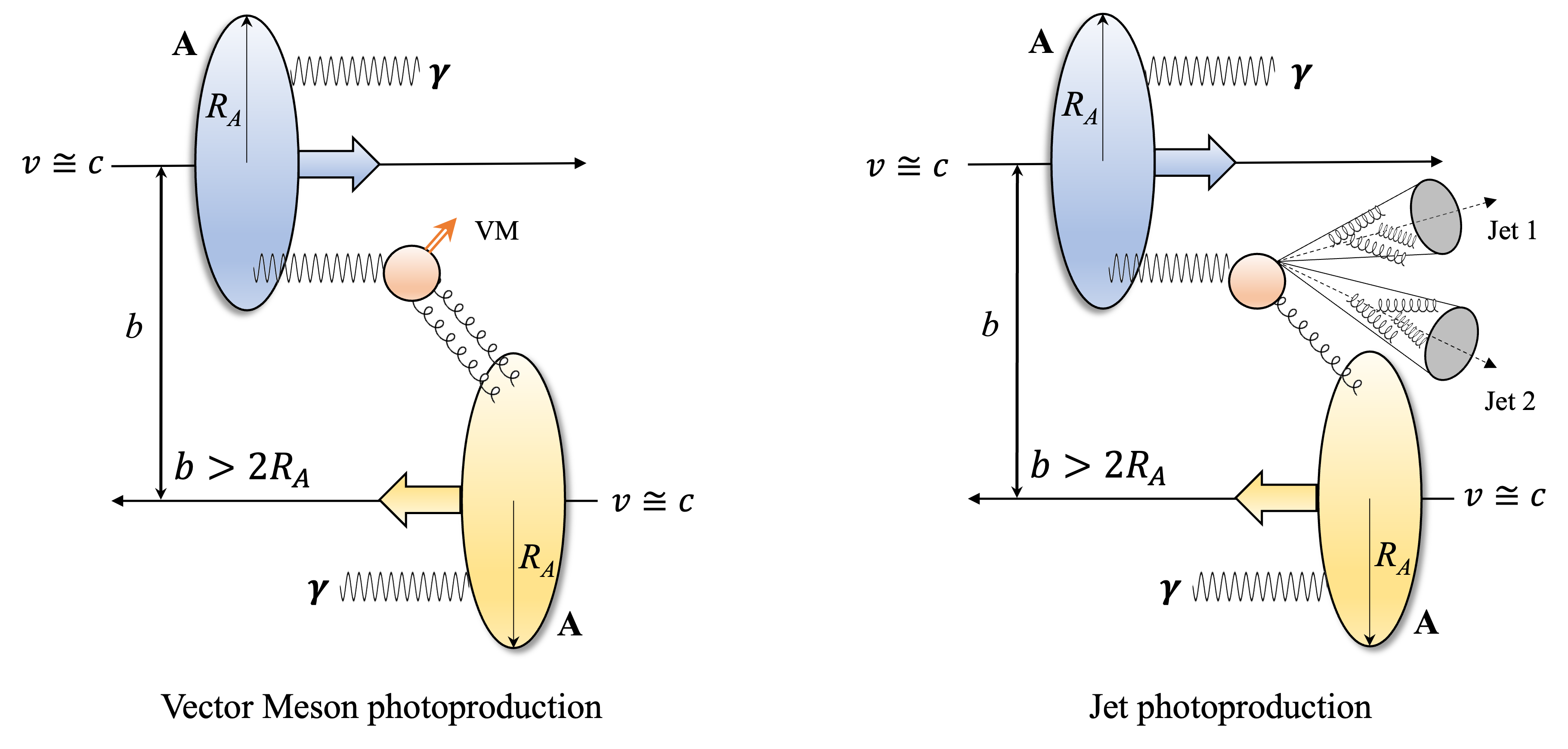}
  \caption{ \label{fig:figure_1} Illustration of UPC Vector Meson (VM) photoproduction (left) and jet photoproduction (right) in heavy-ion A$+$A collisions.}
\end{figure}

\begin{itemize}
    \item \textbf{Virtuality.} Due to the heavy nuclei, photons are emitted with small transverse momentum that has a typical value of $\sim 20-40~\rm{MeV/c}$; the virtuality, $Q^2$, is around $10^{-4}-10^{-3}~\rm{GeV^2}$. The hard scale of the process is usually determined by the final-state particles, e.g., their mass or transverse momentum.
    \item \textbf{Rates.} For QED processes, e.g., two-photon scatterings cross section, the rate of the A+A reaction scales with the nuclear electric charge as $\sim Z^{4}$. (Here $Z$ is the charge of the nucleus, in the units of electron charge $|e|$.) For QCD processes, e.g., photon-Pomeron interactions, the rate scales as $\sim Z^{2}$.
    \item \textbf{Energy reach.} The center-of-mass energy between the photon and per nucleon in the nucleus, $W_{\gamma*N}$, ranges from less than 10 GeV (RHIC) to $\sim 1~\rm{TeV}$ (LHC). This corresponds to momentum fraction $x$ in the range from $10^{-1}$ to $10^{-5}$. Note that, since the saturation scale grows as $Q^{2}_{s} \sim (A/x)^{1/3}$, the UPCs at LHC can probe larger $Q^{2}_{s}$ values than those to be probed in the collisions at the future EIC.
    \item \textbf{Observables.} Most UPCs measurements are for the exclusive and diffractive Vector-Meson (VM) photoproduction. Event kinematics and topology are easy to reconstruct. Recent measurements show that semi-inclusive processes, e.g., inclusive and diffractive jets~\cite{ATLAS:2017kwa}, and inclusive processes of measuring charged particles only~\cite{ATLAS:2021jhn}, are feasible. 
    
\end{itemize}

There are two major heavy-ion collider facilities currently studying UPCs, RHIC and the LHC. With the heavy-ion runs scheduled between 2023 and 2025 both at RHIC and the LHC, there are a couple of experimental opportunities to the UPCs. For example, forward detectors have been added to the STAR experiment, where particle acceptance between 2.5 to 4 in pseudorapidity $\eta$ became available. Typically, the forward rapidity corresponds to high-$x$ kinematic region. In order to unambiguously discover saturation, the high-$x$ region where we do not expect saturation should also be studied on the same footing as the low-$x$ region. In addition, exploration of the large-$x$ region may shine new light to the anti-shadowing region of nuclear PDFs. Currently, resolving the photon energy ambiguity using the forward neutron emissions has been shown possible, extending the kinematic reach at low-$x$ by two orders of magnitude at the LHC.

%
\section{Double ratio in UPCs coming from saturation physics}
%

\subsection{Elastic vector meson production in UPCs} 
\label{sec:elastic}

\begin{figure}[thb]
    \centering
    \includegraphics[width=0.7 \textwidth]{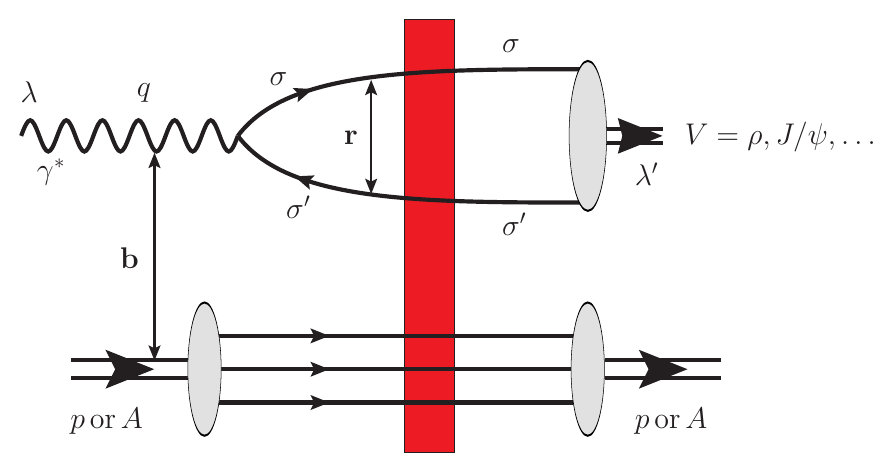}
    \caption{Exclusive vector meson production amplitude in DIS and UPCs at high energy.}
    \label{vmeson amplitude}
\end{figure}

As described above, UPCs (in A$+$A or $p+$A) involve the interaction of a photon of a very small virtuality originating from one nucleus with the other nucleus or proton. At high energy (small $x$), the dominating process involves the photon fluctuating into a $q \bar q$ pair that interacts with the nucleus. One of the possible outcomes of this interaction is the exclusive production of vector meson. The scattering amplitude for exclusive vector meson production in UPCs at leading order is given by the diagram shown in \fig{vmeson amplitude},
where the oval on the upper right represents the vector meson light-cone wave function \cite{Lepage:1980fj}, the ovals on the lower left and lower right represent the target proton or nucleus which remains intact in the interaction, and the vertical rectangle denotes the (mainly gluon) shock wave responsible for the interaction. For the target to remain intact, the shock wave should transfer no color away from the target, i.e., the exchange needs to be color-singlet. To obtain the cross section, we need to square the scattering amplitude and integrate over the phase space of the final state particles. For the convenience of our later calculation, we should Fourier transform the amplitude into transverse coordinate space. The result for the elastic vector meson production cross section at leading order is \cite{Ryskin:1992ui, Brodsky:1994kf} (see \cite{Kovchegov:2012mbw} for a pedagogical derivation in the transverse coordinate space)
\begin{equation}\label{eq:diffraction}
    \sigma^{\gamma^{*} {\rm A} \to V {\rm A}}_\textrm{el} =\frac{1}{2}\sum_{\lambda}\sum_{\lambda^{\prime}}\int d^{2} b_\perp\left|\int\frac{d^{2} r_\perp }{4\pi}\int\limits_{0}^{1}\frac{dz}{z(1-z)}\Psi_T^{\gamma^{*}\to q\bar{q}}(\mathbf{r},z) \, N(\mathbf{r},\mathbf{b},Y) \, \Psi^{V}_T(\mathbf{r},z)^{*}\right|^{2}.
\end{equation}
Here $\Psi_T^{\gamma^{*}\to q\bar{q}} (\mathbf{r},z)$ is the light-cone wave function for a transversely polarized photon splitting into a $q\bar{q}$ pair of the transverse separation $\mathbf r$ with the quark carrying a fraction $z$ of the photon's longitudinal momentum, $\Psi_T^{V}$ is the transversely polarized vector meson wave function, and $N(\mathbf{r},\mathbf{b},Y)$ is the dipole amplitude for a $q\bar{q}$ pair interacting with the nucleus at impact parameter $\mathbf{b}$ and rapidity $Y = \ln (s \, r_\perp^2) \approx \ln (1/x)$ introduced above. We use the definition of the light-cone wave functions normalized as in \cite{Kovchegov:2012mbw}. As shown in \fig{vmeson amplitude}, the photon comes in with polarization $\lambda$, while the vector meson is produced with the polarization $\lambda'$. The former is averaged over, assuming that the photon in UPCs is real, due to its very low $Q^2$: therefore, we keep only the transverse polarizations for the photon. This, in turn, implies that the outgoing vector meson has to be transversely polarized as well. We sum over the vector meson's transverse polarizations in \eq{eq:diffraction}. In addition, a sum over quark polarizations $\sigma, \sigma'$ (as denoted in \fig{vmeson amplitude}) and quark colors is implied inside the absolute value brackets in \eq{eq:diffraction}.

In the saturation/CGC framework, the dipole amplitude $N$ is found by using the BK/JIMWLK nonlinear evolution equations
\cite{Balitsky:1995ub,Balitsky:1998ya,Kovchegov:1999yj,Kovchegov:1999ua, Jalilian-Marian:1997dw,Jalilian-Marian:1997gr,Weigert:2000gi,Iancu:2001ad,Iancu:2000hn,Ferreiro:2001qy} with the GGM/MV quasi-classical initial conditions \cite{Mueller:1989st,McLerran:1993ni,McLerran:1993ka,McLerran:1994vd}. Below, we will consider the quasi-classical and small-$x$ evolution regimes separately, analysing $R_{\rm UPC}$ from \eq{eq3} in each case. To proceed with the calculation, we need to delve into a discussion about the photon and vector meson wave functions.

\begin{figure}[h]
    \centering
    \includegraphics[width = 0.5 \textwidth]{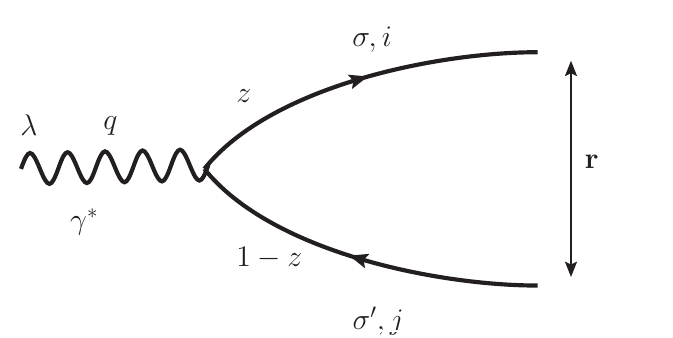}
    \caption{Virtual photon wave function at leading order.}
    \label{fig:virtual photon wave function}
\end{figure}

The virtual photon wave function at the leading order is well known \cite{Bjorken:1970ah, Nikolaev:1990ja}. The corresponding diagram for a photon splitting into a quark-antiquark pair is shown below in \fig{fig:virtual photon wave function}. In the case of UPCs, the photon has a very small $Q^{2}$, so the longitudinally polarized photon wave function is small compared to the transversely polarized one: as already mentioned above, we will ignore the former in our calculation. 
The transverse coordinate space transversely-polarized photon wave function that appears in \eq{eq:diffraction} is
\begin{equation}\label{eq7}
    \Psi_{T}^{\gamma^{*}\to q\bar{q}}(\mathbf{r},z)=\frac{eZ_{f}}{2\pi}\sqrt{z(1-z)} \, \delta_{ij}\left[-i \, \delta_{\sigma, -\sigma^{\prime}} \, (1-2z-\sigma\lambda)\, \frac{\mathbf{\epsilon}^{\lambda} \cdot \mathbf{r}}{r_{\perp}} \, \partial_{r_{\perp}}+\delta_{\sigma\sigma^{\prime}} \, \frac{m_{f}}{\sqrt{2}} \, (1+\sigma\lambda)\right] K_{0}(r_{\perp}a_{f}),
\end{equation}
where $\sigma,\sigma'$ are the polarizations and $i,j$ are the
colors of the quark and antiquark, respectively. We have defined $a_{f}^{2} \equiv z(1-z) \, Q^{2} +m_{f}^{2}$ with $m_f$ the mass of quarks with flavor $f$. The magnitude of the electron charge is labelled by $e$, while $Z_f$ is the fractional charge of the quark. In addition, $\pd_{r_{\perp}} = \pd/\pd r_\perp$. As above, the photon has polarization $\lambda$ and $z$ is the longitudinal fraction of photon's momentum carried by the quark.

The vector meson wave function, on the other hand, is more complicated because of its nonperturbative nature. We could, however, obtain a reasonable form of the vector meson function if we are willing to make a few assumptions. For our purposes we will follow the approach put forth in \cite{Kowalski:2003hm, Nemchik:1994fp, Nemchik:1996cw, Dosch:1996ss, Forshaw:2003ki, Kowalski:2006hc}. The starting point of this approach is the assumption that the vector meson is predominantly a quark--antiquark state that has the same polarization structure as a virtual photon. By making the replacement
\begin{equation}\label{eq8}
    \frac{eZ_{f}}{2 \pi} z(1-z) K_{0}(r_{\perp}a_{f}) \rightarrow\phi_{T}(r_{\perp},z),
\end{equation}
for the scalar part of the wave function, we can write down the transversely polarized vector meson wave function as
\begin{equation}\label{eq9}
    \Psi_{T}^{V}(\mathbf{r},z)=\frac{1}{\sqrt{z(1-z)}}\delta_{ij}\left[ -i \, \delta_{\sigma, - \sigma^{\prime}} \, (1-2z-\sigma\lambda)\frac{\mathbf{\epsilon}^{\lambda}\cdot\mathbf{r}}{r_{\perp}}\partial_{r_{\perp}}+\delta_{\sigma\sigma^{\prime}}\frac{m_f}{\sqrt{2}}(1+\sigma\lambda)\right]\phi_{T}(r_{\perp},z).
\end{equation}
Note again that transversely polarized photon produces only transversely polarized vector meson;  
therefore, in UPCs transverse vector mesons production is dominant (with the longitudinal vector meson production being suppressed by a factor of $Q^{2}/Q_s^2 \ll 1$). We will employ only the transversely polarized vector meson wave function. 

The scalar part of the vector meson wave function can be reasonably described using a ``boosted Gaussian" form \cite{Brodsky:1980vj, Nemchik:1994fp, Nemchik:1996cw, Forshaw:2003ki}. We start by assuming that in the rest frame of the vector meson, the wave function is Gaussian in the three-momentum of the quark and antiquark, so that
\begin{equation}\label{eq10}
    \tilde{\phi}_{T}(\vec{p} , z)\propto\exp(-\frac{R^{2}}{2}\Vec{p}^{\, 2}),
\end{equation}
where $\vec{p}$ is the three-momentum of the quark (or antiquark) and $R$ is the typical radius of a vector meson. The momentum $\vec{p}$ is related to the invariant mass through
\begin{equation}\label{eq11}M^{2}=4E^{2}=4(\vec{p}^{\, 2}+m_{f}^{2}).
\end{equation}
In the boosted frame, we have a fast moving vector meson with its quark and antiquark having longitudinal momentum fractions $z$ and $1-z$, respectively. The four momenta of the on-mass-shell quark and antiquark are
\begin{align}\label{eq12}
    p_{1}^{\prime}&=\left( zP^{+},\frac{k_{T}^{2}+m_{f}^{2}}{zP^{+}},\mathbf{k} \right),\nonumber\\
    p_{2}^{\prime}&= \left( (1-z)P^{+},\frac{k_{T}^{2}+m_{f}^{2}}{(1-z)P^{+}},-\mathbf{k} \right),
\end{align}
where $\mathbf{k}$ is the transverse momentum of the quark (or antiquark) in the boosted frame and we use the spacetime metric $g_{+-}=g_{-+}=\frac{1}{2}$, $g_{11}=g_{22}=-1$. We assume that the meson is moving in the light-cone plus direction with a large momentum $P^+$. The invariant mass is 
\begin{equation}\label{eq13}
    M^{2}= (p'_1 + p'_2)^2
    =\frac{k_{T}^{2}+m_{f}^{2}}{z(1-z)}
\end{equation}
with $k_T = |\mathbf{k}|$. Equating the invariant masses in two frames, Eqs.~\eqref{eq11} and \eqref{eq13}, we can relate $\vec{p}^{\, 2}$ and $k_{T}^2$, 
\begin{equation}\label{eq14}
    \vec{p}^{\, 2}=\frac{k_{T}^{2}+m_{f}^{2}}{4z(1-z)}-m_{f}^{2}.
\end{equation}
Further, if we assume that the wave function is boost-invariant (i.e., the scalar part of the wave function takes on the same value after a Lorentz transformation), then in the boosted frame the wave function has the form (cf. \eq{eq10})
\begin{equation}\label{eq15}
    \tilde{\phi}_{T}(k_{T},z)\propto\exp\left[-\frac{R^{2}}{8} \left( \frac{k_{T}^{2}+m_{f}^{2}}{z(1-z)}-4m_{f}^{2} \right) \right].
\end{equation}

Fourier-transforming the wave function into the transverse coordinate space, we have
\begin{equation}\label{eq16}
    \phi_{T}(r_{\perp},z)\propto z(1-z)\exp\left[-\frac{2z(1-z) \, r_{\perp}^{2}}{R^{2}}-\frac{m_{f}^{2}R^{2}}{8z(1-z)}+\frac{m_{f}^{2}R^{2}}{2}\right].
\end{equation}
Substituting the result \eqref{eq16} into \eq{eq9}, along with the normalization factor ${\cal N}_T$, we obtain an expression for the transversely polarized vector meson wave function \cite{Kowalski:2003hm, Nemchik:1994fp, Nemchik:1996cw, Dosch:1996ss, Forshaw:2003ki},
\begin{align}\label{eq17}
    \Psi_T^{V}(r_{\perp},z)= {\cal N}_{T}\sqrt{z(1-z)} \, \delta_{ij}\left[i \, \delta_{\sigma, -\sigma^{\prime}} \, (1-2z-\sigma\lambda)\frac{4z(1-z)}{R^{2}} \, \mathbf{\epsilon}^\lambda \cdot \mathbf{r}\right.\\
    +\left.\delta_{\sigma\sigma^{\prime}}\frac{m_{f}}{\sqrt{2}}(1+\sigma\lambda)\right]\exp\left[-\frac{2z(1-z) \, r_{\perp}^{2}}{R^{2}}-\frac{m_{f}^{2}R^{2}}{8z(1-z)}+\frac{m_{f}^{2}R^{2}}{2}\right]. \notag
\end{align}
The wave function \eqref{eq17} is normalized such that 
\begin{align}\label{norm}
    \sum_{\sigma, \sigma', i, j} \, \int\limits_0^1 \frac{dz}{z (1-z)} \, \int \frac{d^2 r_\perp}{4 \pi} \, \left| \Psi_T^{V}(r_{\perp},z) \right|^2 = 1, 
\end{align}
resulting from the assumption that the vector meson is predominantly a quark--antiquark pair (in all frames). The relation \eqref{norm} can be used to find ${\cal N}_T$ once the vector meson radius $R$ is found. The procedure for finding $R$ is described in \cite{Kowalski:2006hc} and involves fitting the coupling of the vector meson to the electromagnetic current $f_V$. Since the Particle Data Group (PDG) \cite{Workman:2022ynf} value of the charm quark mass along with the values of the $\rho$ meson mass and $f_V$ for $J/\psi$ are somewhat different now compared to the numbers used in \cite{Kowalski:2006hc}, we had to re-evaluate ${\cal N}_T$ and $R$ for both $\rho$ and $J/\psi$ mesons.
The resulting parameters for the wave function \eqref{eq17} for the $\rho$ and $J/\psi$ mesons are listed in Table~\ref{tab:parameters} below. 

Before we proceed, we note here that vector meson light-cone wave functions are, indeed, model-dependent. It would therefore be hard to estimate the accuracy of a specific numerical prediction based on those wave functions. However, our goal in this work is more qualitative than quantitative. For our purposes, we need the $J/\psi$ wave function to be dominated by smaller $q \bar q$ pairs, while the $\rho$-meson wave function needs to favor larger dipole sizes. The wave function \eqref{eq17} certainly satisfies these properties.

\begin{figure}[thb]
    \centering
    \includegraphics[width=0.7\textwidth]{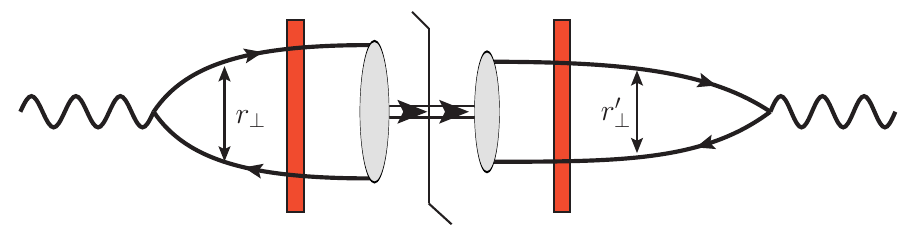}
    \caption{A diagrammatic representation of \eq{eq:elastic vmp}. Vertical rectangles denote the shock waves in the amplitude and in the complex conjugate amplitude, while the solid vertical line denotes the final state cut.}
    \label{fig:5}
\end{figure}

For future calculations, it is easier to simplify \eq{eq:diffraction} a little. Substituting Eqs.~(\ref{eq7}) and (\ref{eq17}) into Eq.~(\ref{eq:diffraction}), summing over polarizations $\lambda, \lambda', \sigma, \sigma'$ and quark colors, and integrating over the angles of $\mathbf{r}$ while assuming the dipole amplitude to be independent of the $\mathbf{r}$ direction, we arrive at 

\begin{align}\label{eq:elastic vmp}
    \sigma^{\gamma^{*} {\rm A} \to V {\rm A}}_\textrm{el} = & \, \frac{\alpha_{EM}Z_{f}^{2}}{\pi} \, N_{c}^{2} \, {\cal N}_{T}^{2}\,e^{m_{f}^{2}R^{2}}\int\limits_0^1 dz \, dz' \int\limits_0^\infty dr_{\perp} \, dr'_{\perp} \, r_{\perp} \, r_{\perp}'\\ &\times\left\lbrace\frac{4}{R^{2}}z(1-z) \left[z^2 + (1-z)^{2} \right] \, r_{\perp}\, a_{f} \, K_{1}(r_{\perp}a_{f})+m_{f}^{2} \, K_{0}(r_{\perp}a_{f})\right\rbrace \, \exp\left[-\frac{2z(1-z) \, r_{\perp}^{2}}{R^{2}}-\frac{m_{f}^{2}R^{2}}{8z(1-z)}\right]\nonumber\\ 
    &\times\left\lbrace\frac{4}{R^{2}}z'(1-z') \left[z^{\prime \, 2} + (1-z')^{2} \right] \, r_{\perp}' \, a_{f}' \, K_{1}(r_{\perp}'a_{f}')+m_{f}^{2}K_{0}(r_{\perp}'a_{f}')\right\rbrace\exp\left[-\frac{2z'(1-z')r_{\perp}'^{2}}{R^{2}}-\frac{m_{f}^{2}R^{2}}{8z'(1-z')}\right]\nonumber\\ 
    & 
    \times\int d^{2}b_\perp \, N(r_\perp ,b_\perp , Y) \, N(r'_\perp, b_\perp, Y). \nonumber
\end{align}
Here $a_f^{\prime \, 2} \equiv z' (1-z') \, Q^2 + m_f^2$ and $\alpha_{EM} = e^2/(4 \pi)$. The assumption of $N(\mathbf{r}, \mathbf{b}, Y) = N(r_\perp ,b_\perp , Y)$ that we employ here is valid in the quasi-classical approximation for a large nucleus \cite{Mueller:1989st,McLerran:1993ni,McLerran:1993ka,McLerran:1994vd} and for the leading contribution to high-energy scattering (see, e.g., \cite{Kovchegov:2012mbw}). 
Diagrammatically, \eq{eq:elastic vmp} can be represented by the graph in \fig{fig:5}, where we separately integrate over the transverse sizes $r_{\perp}, r_{\perp}'$ of the dipole in the amplitude and in the complex conjugate amplitude, and over the impact parameter $\mathbf{b}$, common to both amplitudes. 



\subsection{Inclusive hadron or jet production}
\label{sec:inclusive}

The ratio $R_{\rm{UPC}}$ in \eq{eq3} also contains the inclusive hadron or jet production cross section. Our goal here is to assess the behavior of $R_{\rm{UPC}}$ with varying nuclear atomic number $A$. We will assume that at small-$x$ fragmentation and jet showering happen mainly outside the cold nuclear medium of the target nucleus and are, therefore, $A$-independent. Hence, we will omit such final-state interactions from our analysis and concentrate on calculating the quark production cross section, which we will then use to derive the $A$-dependence of the double ratio $R_{\rm{UPC}}$. 


\begin{figure}[thb]
    \centering
    \includegraphics[scale=0.7]{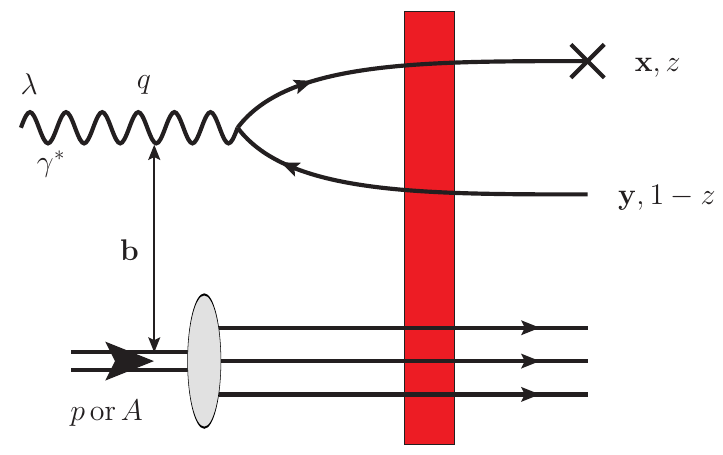}
    \caption{A diagram contributing to the scattering amplitude for the inclusive single quark production in the virtual photon--target proton (nucleus) scattering. The measured quark is marked by a cross.}
    \label{fig:6}
\end{figure}

To calculate the inclusive quark production cross section, we again consider the dipole picture of the photon--nucleus scattering, which is the dominant contribution at high energy (small $x$). A typical contributing diagram is shown in \fig{fig:6}, constructed by analogy to \fig{vmeson amplitude}. The difference now is that the target proton or nucleus can break up, which means the interaction is no longer color-singlet. Moreover, the $q\bar q$ pair resulting from the virtual photon's splitting does not recombine into a vector meson. Both the quark and antiquark in the pair can fragments into independent hadrons or jets. We will calculate the quark production cross section (with the produced quark denoted by the cross in \fig{fig:6}): in the eikonal approximation employed here, the antiquark production cross section is exactly the same. 

The scattering amplitude in \fig{fig:6} can be factorized into a convolution of the photon light cone wave function with the dipole amplitude for the $q\bar{q}$ pair scattering on the nucleus. In a light-cone perturbation theory calculation \cite{Lepage:1980fj}, to the diagram in \fig{fig:6} one has to add the non-interacting contribution where the $\gamma\to q\bar{q}$ splitting takes place to the right of the shock wave. The result for the differential cross section $\frac{d\sigma}{d^{2}p_T}$ was first derived in \cite{Mueller:1999wm} and is also given by Eq.~(85) of \cite{Kovchegov:2015zha} in terms of the dipole $S$-matrix. Recalling that this $S$-matrix is $S = 1-N$ for the dipole amplitude $N$, we write
\begin{align}\label{eq33}
    \frac{d\sigma}{d^{2}p_T} =\frac{1}{2(2\pi)^3} & \ \int\limits_0^1 \frac{dz}{z(1-z)} \, \int d^{2}x_{\perp}d^{2}x_{\perp}^{\prime}d^{2}y_{\perp} \, e^{-i\mathbf{p}\cdot(\mathbf{x}-\mathbf{x}^{\prime})} \, \frac{1}{2} \, \sum_\lambda\Psi^{\gamma\to q\bar{q}}_T (\mathbf{x}-\mathbf{y}, z) \, \left[\Psi_T^{\gamma\to q\bar{q}} (\mathbf{x}^{\prime}-\mathbf{y}, z) \right]^{*}  \\ 
    &  \times \, \left[ N \left( \mathbf{x}-\mathbf{y} , \frac{\mathbf{x}+\mathbf{y}}{2} , Y \right) + N \left( \mathbf{x}'-\mathbf{y} , \frac{\mathbf{x}'+\mathbf{y}}{2} , Y \right)  -N \left( \mathbf{x}-\mathbf{x}' , \frac{\mathbf{x}+\mathbf{x}'}{2} , Y \right) \right] , \nonumber
\end{align}
where the sum over the final-state quark polarizations and colors is implied in the product of light-cone wave functions.
As we have mentioned earlier, in UPCs the virtual photon has a very small virtuality ($Q^{2}\approx 400-900~\text{MeV}^{2}$). Therefore we only need to consider transversely polarized photons, with \eq{eq33} including the averaging over the virtual photon polarizations. Also, note that we are continuing to use the $N (\mathbf{r}, \mathbf{b}, Y)$ notation for the dipole amplitudes, resulting in a slightly more complicated than usual arguments of the dipole amplitudes in \eq{eq33}.

The expression \eqref{eq33} can also be further simplified. Substituting the virtual photon light-cone wave function from \eq{eq7} into \eq{eq33}, averaging over photon polarizations and summing over final-state quark and antiquark polarizations, we obtain
\begin{align}\label{eq34}
&  \frac{d\sigma}{d^{2}p_T}=\frac{2N_c \, \alpha_{EM}Z_{f}^{2}}{(2\pi)^4} \int d^{2}x_{\perp}d^{2}x_{\perp}^{\prime}d^{2}y_{\perp}e^{-i\mathbf{p}\cdot(\mathbf{x}-\mathbf{x}^{\prime})}  \int\limits_0^1 dz \\
& \times \, \left\lbrace\left[z^{2}+(1-z)^{2}\right]a_{f}^{2}\frac{(\mathbf{x}-\mathbf{y})\cdot(\mathbf{x}^{\prime}-\mathbf{y})}{|\mathbf{x}-\mathbf{y}||\mathbf{x}^{\prime}-\mathbf{y}|}K_{1}(|\mathbf{x}-\mathbf{y}|a_{f}) \,  K_{1}(|\mathbf{x}^{\prime}-\mathbf{y}|a_{f})+m_{f}^{2}K_{0}(|\mathbf{x}-\mathbf{y}|a_{f})K_{0}(|\mathbf{x}^{\prime}-\mathbf{y}|a_{f})\right\rbrace \notag \\ 
& \times \, \left[ N \left( \mathbf{x}-\mathbf{y} , \frac{\mathbf{x}+\mathbf{y}}{2} , Y \right) + N \left( \mathbf{x}'-\mathbf{y} , \frac{\mathbf{x}'+\mathbf{y}}{2} , Y \right)  -N \left( \mathbf{x}-\mathbf{x}' , \frac{\mathbf{x}+\mathbf{x}'}{2} , Y \right) \right] . \notag
\end{align}
Some of the integrals in \eq{eq34} can be carried out explicitly. This is done in Appendix~\ref{A}, resulting in
\begin{align}\label{eq:inclusive}
    &\frac{d\sigma}{d^{2}p_T}=\frac{N_{c} \, \alpha_{EM}Z_{f}^{2}}{(2\pi)^{3}} \int\limits_0^1 dz \, \int d^{2}r_{\perp}d^{2}b_{\perp}e^{-i\mathbf{p}\cdot\mathbf{r}}\left\lbrace[z^{2}+(1-z)^{2}]\left[4ia_{f}K_{1}(r_{\perp}a_{f})\frac{\mathbf{p}\cdot\mathbf{r}}{(p_{\perp}^{2}+a_{f}^{2})r_{\perp}}-2 \, K_{0}(r_{\perp}a_{f})\right.\right. \\ 
    & \left. + \, r_{\perp} \, a_{f} \, K_{1}(r_{\perp}a_{f}) \Bigg]+\frac{4m_{f}^{2}}{p_{\perp}^{2}+a_{f}^{2}}K_{0}(r_{\perp}a_{f})-\frac{m_{f}^{2}r_{\perp}}{a_{f}}K_{1}(r_{\perp}a_{f})\right\rbrace N(\mathbf{r},\mathbf{b}, Y). \notag
\end{align}
In arriving at \eq{eq:inclusive}, we have employed the fact that
\begin{align}\label{tr_symm}
    \int d^2 b_\perp \, N (\mathbf{r},\mathbf{b}, Y) = \int d^2 b_\perp \, N (- \mathbf{r},\mathbf{b}, Y),
\end{align}
which is valid due to $N$ being a scalar quantity and our unpolarized proton or nuclear target having no preferred transverse direction. 

%
\section{Estimates of the double ratio in UPCs}
%
\label{sec:estimates}

Our next goal is to study the dependence of the double ratio $R_{\rm{UPC}}$ from \eq{eq3} on the atomic number $A$ for different vector mesons. We will pick the $\rho$ meson as an example of a vector meson whose wave function is dominated by ``large" dipole sizes, and $J/\psi$ meson as an example of a small-dipole-size meson. Below we evaluate the $A$-dependence of $R_{\rm{UPC}}$ for these two mesons analytically and  numerically. We will first work in the quasi-classical GGM/MV approximation \cite{Mueller:1989st,McLerran:1993ni,McLerran:1993ka,McLerran:1994vd}, then we will include LO small-$x$ BK/JIMWLK evolution \cite{Balitsky:1995ub,Balitsky:1998ya,Kovchegov:1999yj,Kovchegov:1999ua,Jalilian-Marian:1997dw,Jalilian-Marian:1997gr,Weigert:2000gi,Iancu:2001ad,Iancu:2000hn,Ferreiro:2001qy}.


\subsection{Double ratio in the quasi-classical approximation}

\label{sec:classical}


\subsubsection{Elastic $J/\psi$ and $\rho$ production: heuristic estimates and numerical integration}\label{classical vm scaling}

In \eq{eq:diffraction} (or \eq{eq:elastic vmp}), the dipole amplitude $N(\mathbf{r},b,Y)$ describes the interaction of the quark-antiquark pair with the target nucleus. In processes where saturation effects are taken into account, one has to include multiple gluon exchanges 
between the $q \bar q$ pair and the nucleus. Including $t$-channel gluon exchanges to all orders in the GGM/MV model leads to the dipole amplitude \cite{Mueller:1989st}
\begin{equation}\label{GGM/MV}
    N(\mathbf{r},\mathbf{b}, Y)=1-\exp\left\lbrace -\frac{r_{\perp}^{2}Q_{s}^{2}(\mathbf{b})}{4}\ln\frac{1}{r_{\perp}\Lambda}\right\rbrace
\end{equation}
with the saturation scale $Q_{s}$ given by
\begin{align}
    Q_s^2 (\mathbf{b}) = 4 \pi \as^2 \, \frac{C_F}{N_c} \, T (\mathbf{b}).
\end{align}
Here $T (\mathbf{b})$ is the nuclear profile (thickness) function, 
\begin{align}\label{Tb}
    T (\mathbf{b}) \equiv \int\limits_{-\infty}^\infty dz \, \rho (\mathbf{b}, z) ,
\end{align}
where $\rho (\mathbf{b}, z)$ is the nucleon number density in the nucleus, $\Lambda$ is the infrared (IR) cutoff, and $C_F$ is the fundamental Casimir operator of SU($N_c$).
We should point out that the dipole amplitude given in \eq{GGM/MV} does not include small-$x$ evolution: this is why $Q_s^2 (\mathbf{b})$ here is independent of energy/rapidity $Y$, leading to similarly energy-independent dipole amplitude $N(\mathbf{r},\mathbf{b}, Y)$ in \eq{GGM/MV}.  

Our goal now is to determine the dependence of the elastic VM production cross section on the atomic number $A$. After a closer inspection of \eq{eq:elastic vmp}, we see that the $A$-dependence is contained entirely in the $\mathbf{b}$-integral 
\begin{align}
    \int d^{2}b_\perp \, N(r_\perp ,b_\perp , Y) \, N(r'_\perp, b_\perp, Y)
\end{align}
over the transverse area of the nucleus. 
This integral is hard to evaluate exactly analytically. Therefore, we have to make approximations for the dipole amplitude $N(\mathbf{r},\mathbf{b}, Y)$ based on whether $r_{\perp}$ and $r'_\perp$  are larger or smaller than $1/Q_{s}(\mathbf{b})$, which corresponds to the dipole $r_\perp$ and/or the dipole $r'_\perp$ being inside or outside the saturation regime (see \fig{fig:5}). Since the integrations over $r_{\perp}$ and $r_{\perp}'$ range over all positive values between $0$ and $\infty$, we have three cases to consider: (i) $r_{\perp},r_{\perp}' \ll 1/Q_{s}$, (ii) $r_{\perp},r_{\perp}' \gtrsim 1/Q_{s}$, and (iii) $r_{\perp}\ll 1/Q_{s},r_{\perp}' \gtrsim 1/Q_{s}$. The case when $r'_{\perp}\ll 1/Q_{s}, r_{\perp} \gtrsim 1/Q_{s}$ gives the same contribution as the case (iii), due to the $\mathbf{r} \leftrightarrow \mathbf{r}'$ symmetry of \eq{eq:elastic vmp}. As follows from \eq{eq:elastic vmp}, the dipole sized $r_\perp$ and $r'_\perp$ are controlled by the convolutions of the virtual photon and vector meson wave functions with the dipole size dependence of the amplitude $N$.

In these three regions we obtain different $A$-scaling, using the following arguments:
\begin{enumerate}[(\romannumeral 1)]
\item $r_{\perp},r_{\perp}' \ll 1/Q_{s}$: We approximate the dipole amplitude \eqref{GGM/MV} outside the saturation region by expanding it to the lowest order in $r_\perp \, Q_{s}(\mathbf{b})$, such that 
    \begin{subequations}\label{NNout}
    \begin{align}\label{eq20}
      &N(\mathbf{r},\mathbf{b},Y) \bigg|_{r_\perp \, Q_{s}(\mathbf{b}) \ll 1} \approx\frac{r_{\perp}^{2}Q_{s}^{2}(\mathbf{b})}{4}\ln\frac{1}{r_{\perp}\Lambda} \propto A^{1/3},  \\
      &N(\mathbf{r}',\mathbf{b},Y)\bigg|_{r'_\perp \, Q_{s}(\mathbf{b}) \ll 1} \approx\frac{r_{\perp}'^{2}Q_{s}^{2}(\mathbf{b})}{4}\ln\frac{1}{r_{\perp}'\Lambda} \propto A^{1/3}, 
    \end{align}        
    \end{subequations}
where the last proportionality follows from $Q_{s}^{2} (\mathbf{b}) \propto T (\mathbf{b}) \propto A^{1/3}$. Since the area integral scales as $\int d^2 b_\perp \sim A^{2/3}$, we conclude that
    \begin{equation}\label{eq21}
        \int d^{2} b_\perp \, N(\mathbf{r},\mathbf{b},Y) \, N(\mathbf{r}',\mathbf{b},Y) \bigg|_{r_{\perp},r_{\perp}' \ll 1/Q_{s} } \propto A^{4/3} .
    \end{equation}
\item $r_{\perp},r_{\perp}' \gtrsim 1/Q_{s}$: Inside the saturation region we approximate 
    \begin{equation}\label{eq22}
        N(\mathbf{r},\mathbf{b},Y)\bigg|_{r_\perp \, Q_{s}(\mathbf{b}) \gtrsim 1} \approx N(\mathbf{r}',\mathbf{b},Y)\bigg|_{r'_\perp \, Q_{s}(\mathbf{b}) \gtrsim 1} \approx 1 ,
    \end{equation}
    such that 
    \begin{equation}\label{eq23}
        \int d^{2} b_\perp \, N(\mathbf{r},\mathbf{b},Y) \, N(\mathbf{r}',\mathbf{b},Y)\bigg|_{r_{\perp},r_{\perp}' \gtrsim 1/Q_{s} } \propto A^{2/3} . 
    \end{equation}
    \item $r_{\perp}\ll 1/Q_{s},r_{\perp}' \gtrsim 1/Q_{s}$ (or $r'_{\perp}\ll 1/Q_{s},r_{\perp} \gtrsim 1/Q_{s}$): With one dipole being outside the saturation region, and another one being inside, we have  
    \begin{subequations}
    \begin{align}\label{eq24}
        &N(\mathbf{r},\mathbf{b},Y)\bigg|_{r_\perp \, Q_{s}(\mathbf{b}) \ll 1} \approx\frac{r_{\perp}^{2}Q_{s}^{2}}{4}\ln\frac{1}{r_{\perp}\Lambda}, \\
        & N(\mathbf{r}',\mathbf{b},Y)\bigg|_{r'_\perp \, Q_{s}(\mathbf{b}) \gtrsim 1} \approx 1.
    \end{align}        
    \end{subequations}
    This leads to 
    \begin{equation}\label{eq25}
        \int d^{2}b_\perp \, N(\mathbf{r},\mathbf{b},Y) \, N(\mathbf{r}',\mathbf{b},Y) \bigg|_{r_{\perp}\ll 1/Q_{s},r_{\perp}' \gtrsim 1/Q_{s}} \propto A .
    \end{equation}
\end{enumerate}
Hence, we conclude that the elastic vector meson production cross section scales with $A$ as a power of $A$, 
\begin{align}\label{alpha}
   \sigma_\textrm{el}^{\gamma^{*} {\rm A} \to V {\rm A}}\propto A^\alpha,  
\end{align}
with $\alpha$ between $2/3$ and $4/3$. The precise power of the scaling depends on the size of the vector meson: if the size of the vector meson is small (e.g., $J/\psi$), then the integral contribution would be dominated by region (\romannumeral 1), and $\sigma_\textrm{el}^{\gamma^{*}{\rm A} \to J/\psi \, {\rm A}}\propto A^{4/3}$; if the size of the vector meson is large (e.g., $\rho$), then the integral contribution would be dominated by region (\romannumeral 3), and  $\sigma_\textrm{el}^{\gamma^{*} {\rm A} \to \rho \, {\rm A}}\propto A^{2/3}$. Therefore, a transition from outside the saturation region into the saturation region should lead to the decrease of the (effective) power $\alpha$ defined in \eq{alpha}.

Notice that in \eq{eq:elastic vmp} the integrand as a function of the dipole sizes $r_{\perp}$ and $r_{\perp}'$ is dominated by the Gaussian and the modified Bessel functions (which decrease exponentially at large $r_{\perp}$ and $r_{\perp}'$), so that the main contribution comes from the regions where $r_{\perp},r_{\perp}'<\frac{1}{a_{f}},R$. For $J/\psi$ production in UPCs, where $Q^2 \approx 0$ and $a_{f}\approx m_{c}\approx 1.27\ \text{GeV}$, this corresponds to $r_{\perp},r_{\perp}'<\frac{1}{m_c}\approx 0.79\text{ GeV}^{-1}$. At relatively low $x$ ($x$ between $10^{-3}$ and $10^{-4}$), the typical saturation scale for a gold nucleus ($A=197$) is about $Q_s \approx 1\ \text{GeV}$ (see, e.g.,  Fig.~3.14 in \cite{Accardi:2012qut}). We see that the $r_{\perp},r_{\perp}'$-integrals in \eq{eq:elastic vmp} are dominated by the non-saturated region (i), so that $\sigma^{\gamma^{*}A\to J/\psi \, A}\propto A^{4/3}$. However, 
these integrals do include contributions from larger $r_\perp, r'_\perp$, coming from the saturation region. Therefore, in an exact evaluation of \eq{eq:elastic vmp}, 
one may expect to see an $A$-scaling that is slightly slower than $A^{4/3}$, especially at the largest $A$ when $1/Q_{s}$ starts to become comparable to the size of $J/\psi$ and saturation effects start to settle in. 

\begin{figure}[ht]
    \centering
    \includegraphics[scale=0.88]{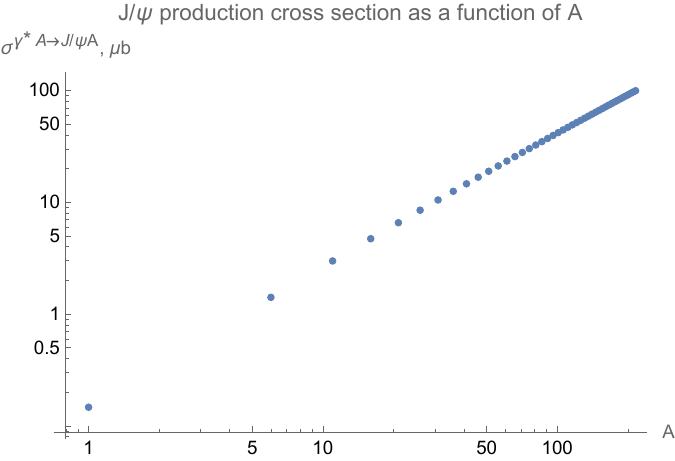}
    \caption{A log-log plot of the elastic $J/\psi$ production cross section versus the atomic number $A$ of the target nucleus. The quasi-classical dipole amplitude used here is given in Eq.(\ref{eq: num dip amp classical}).}
    \label{Jpsi quasi-classical}
\end{figure}

The results of our analysis above are supported by numerical calculations. To evaluate \eq{eq:elastic vmp} numerically for the case of UPCs, we put $Q^2 =900$~MeV$^2$. 
The parameters of the vector meson wave functions, ${\cal N}_{T}$ and $R$, which are used in \eq{eq17} are listed here in Table \ref{tab:parameters}.

\begin{table}[h]
    \centering
    \begin{tabular}{||c c c c|c c||}
        \hline
        Meson & $M_{V}/\rm{GeV}$ & $f_{V}/\rm{GeV}$ & $m_{f}/\rm{GeV}$ & ${\cal N}_{T}$ & ${R}^{2}/\rm{GeV}^{-2}$\\
        \hline
        $J/\psi$ & 3.097 & 0.277 & 1.27 & 0.602 & 2.3\\
        \hline
        $\rho$ & 0.775 & 0.156 & 0.14 & 0.909 & 12.9\\
        \hline
    \end{tabular}
    \caption{Parameters for $J/\psi$ and $\rho$ mesons wave functions given in \eq{eq17} calculated using the prescription outlined in \cite{Kowalski:2006hc} for the current PDG \cite{Workman:2022ynf} values of the quark and meson masses and the coupling $f_V$}.
    \label{tab:parameters}
\end{table}

We use the quasi--classical dipole amplitude that is given by (see, e.g., \cite{Albacete:2009fh}) 
\begin{equation}\label{eq: num dip amp classical}
    N(\mathbf{r},\mathbf{b}, Y)=1-\exp{-\frac{r_{\perp}^{2}Q_{s}^{2}(\mathbf{b})}{4}\ln\left(\frac{1}{r_{\perp}\Lambda}+e\right)},
\end{equation}
to be able to integrate over $r_\perp > 1/\Lambda$ without having an unphysical negative $N$ in that region (cf. \eq{GGM/MV}). For forward kinematics at RHIC we use the value of the saturation scale for Au nuclei at $x \approx 2 \times 10^{-3}$ and $b_\perp =0$ from Fig.~3.9 of \cite{Accardi:2012qut} to write
\begin{align}\label{Qs_cl}
    Q_s^2 (\mathbf{b}) = (1 \, \text{GeV}^2) \, \frac{T(\mathbf{b})}{2 R_A \, \rho } \, \left( \frac{A}{197} \right)^{1/3} \, \theta (R_A - b_\perp)  =  (1 \, \text{GeV}^2) \, \sqrt{1 - \frac{b_\perp^2}{R_A^2}} \, \left( \frac{A}{197} \right)^{1/3} \, \theta (R_A - b_\perp) ,
\end{align}
where, for simplicity, we consider the nucleus to be a sphere of radius $R_A$ with a uniform nucleon number density $\rho=0.40$~fm$^{-3}=0.0031$~GeV$^{3}$, such that $T(\mathbf{b}) = 2 \rho \, \sqrt{R_A^2 - b_\perp^2}$. We approximate $R_A \approx 1.2 A^{1/3}$~fm$= 6 \, A^{1/3}$~GeV$^{-1}$ and choose $\Lambda = 0.2$~GeV for the IR cutoff. We also use $\alpha_{EM} = 1/137$, $\alpha_{s}=1/3$, $N_c =3$, and $Z_f = 2/3$ for $J/\psi$.

In \fig{Jpsi quasi-classical} we present a log-log plot of the $J/\psi$ elastic production cross section versus $A$.  The slope between the two adjacent data points starts at about $\alpha = 1.26$ at $A=1$, then gradually decreases to about $\alpha = 1.12$ at $A=208$ (Pb). 
(The power $\alpha$ is defined in \eq{alpha} above.) We see that the $A$-scaling power $\alpha$ of the elastic $J/\psi$ production in UPCs is smaller than the (analytical) non-saturated value of 4/3, indicating that there are some saturation effects present in the process. The decrease of the slope with increasing $A$ is indeed what we expected, since saturation is more manifest in a larger nucleus. Note also that the magnitude of the cross section $\sigma_\textrm{el}^{\gamma^{*} {\rm A} \to J/\psi \, {\rm A}}$ for Au nuclei, $A=197$, appears to be somewhat above the data recently reported by the STAR collaboration at RHIC~\cite{STAR:2023nos,STAR:2023gpk}, but is closer to the experimental values measured by the ALICE collaboration at the LHC (see Fig.~4 in \cite{ALICE:2023jgu}) for scattering on a similarly-sized Pb nucleus despite the relative crudeness of the approximations we have made here. 




\begin{figure}[h]
    \centering
    \includegraphics[scale=0.8]{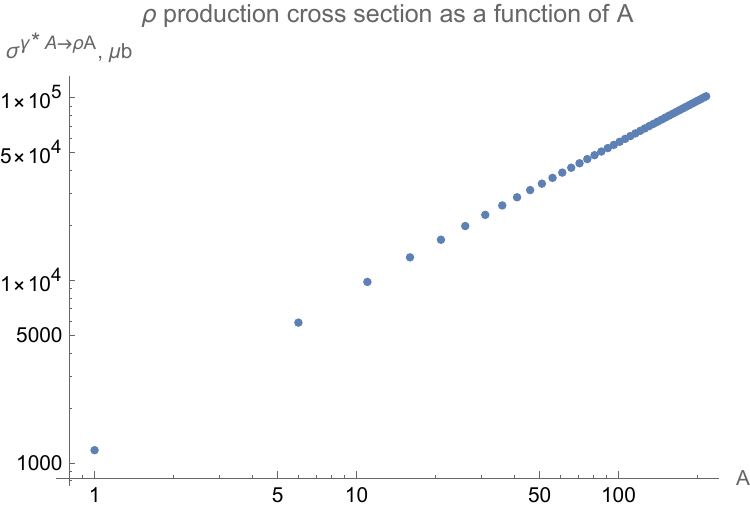}
    \caption{A log-log plot of the elastic $\rho$ production cross section versus the atomic number $A$ of the target nucleus. The quasi-classical dipole amplitude used here is given in Eq.(\ref{eq: num dip amp classical}).}
    \label{rho quasi-classical}
\end{figure}

Similarly, we perform a numerical evaluation of \eq{eq:elastic vmp} for the $\rho$-meson using the same quasi-classical dipole amplitude \eqref{eq: num dip amp classical}. The $\rho$-meson wave function parameters are also given in Table \ref{tab:parameters}. For the $\rho$ meson one uses an effective value of $Z_f = 1/\sqrt{2}$ \cite{Kowalski:2006hc}. In \fig{rho quasi-classical} we show the log-log plot of the elastic $\rho$ production cross section in UPCs versus $A$. The slope between the adjacent data points starts at about $\alpha = 0.90$ at $A=1$, and gradually decreases to about $\alpha = 0.75$ at $A=208$. We see that the $A$-scaling power $\alpha$ is smaller than that for $J/\psi$, as we expected, since $\rho$ is a larger meson which is more affected by saturation effects, which tend to reduce $\alpha$. Moreover, extrapolating the plot in \fig{rho quasi-classical} to the unrealistically high $A=300$, we see that the power $\alpha$ approaches 
the ``ideal" value of 2/3 for a fully saturated process in the regime (ii) considered above, thus agreeing with our analytic estimate.

To summarize our observations here, we see that saturation effects tend to reduce the effective power $\alpha$ from \eq{alpha}. A transition to the saturation regime can happen in two ways. We can increase the atomic number $A$ of the nucleus, while considering the elastic production cross section for the same vector meson. Saturation would lead to the decrease of the slope $\alpha$ with $A$, as observed separately in \fig{Jpsi quasi-classical} and in \fig{rho quasi-classical}. Alternatively, saturation can be reached by increasing the size of the vector meson. Hence, in going from $J/\psi$ to $\rho$ meson, by comparing \fig{Jpsi quasi-classical} to \fig{rho quasi-classical}, we see that $\alpha$ decreases as well between these two figures due to saturation effects.


\subsubsection{
Inclusive production cross section}

Now let us use the above estimates to evaluate the $A$-scaling of the double ratio $R_{\rm{UPC}}$ from \eq{eq3}. Since the elastic cross section has been found above, we also need the $A$ scaling of the inelastic hadron/jet production cross section. To obtain that we will use \eq{eq:inclusive}, in which the $r_\perp$ integral is restricted to $r_\perp < \min \{ 1/p_T, 1/a_f \}$ by the Fourier exponential and by the modified Bessel functions coming from the virtual photon wave function. Since the typical jet or net hadron production cross section is dominated by light flavors, we see that in UPCs $a_f$ for \eq{eq:inclusive} would be rather small, of the order of several tens of MeV, such that $a_f \ll p_T$: hence, we only have the $r_\perp < 1/p_T$ constraint on the range of $r_\perp$, before we account for the dipole amplitude $N$.

For $p_T \gg Q_s$ we have $r_\perp \ll 1/Q_s$ in \eq{eq:inclusive}, placing us outside the saturation region. Using the approximation from Eqs.~\eqref{NNout} for the dipole amplitude, we see that $N \propto A^{1/3}$, such that 
\begin{align}\label{incl_out}
    \frac{d\sigma}{d^{2}p_T} \bigg|_{p_T \gg Q_s} \propto A 
\end{align}
after we integrate over the impact parameters $\mathbf{b}$ between the dipole and the nucleus.

For low momenta, $p_T \lesssim Q_s$ (while still $p_T \gg \Lambda_{QCD}$ with $\Lambda_{QCD}$ the QCD confinement scale), we have the typical $r_\perp \sim 1/p_T \gtrsim 1/Q_s$, such that the $r_\perp$ integral in \eq{eq:inclusive} is largely inside the saturation region. We then apply the approximation $N \approx 1$ in \eq{eq:inclusive} (see \eqref{eq22}) to obtain 
\begin{align}\label{incl_in}
    \frac{d\sigma}{d^{2}p_T} \bigg|_{p_T \lesssim Q_s} \propto A^{2/3} 
\end{align} 
after integration over $\mathbf{b}$. The results in Eqs.~\eqref{incl_out} and \eqref{incl_in} are consistent with the expectations of saturation physics for particle production in dilute-dense scattering, like the proton--nucleus collisions considered in \cite{Kovchegov:1998bi}. Similar to the elastic vector meson production process considered above, saturation effects tend to reduce the power of $A$ for the inclusive production cross section as well. 

\subsubsection{$A$-scaling for the double ratio for $J/\psi$ and $\rho$ production}

Let us first summarize our results for the elastic vector meson production and the inelastic hadron/jet production, before constructing cross section ratios. The elastic vector production cross section scales approximately as 
\begin{align}
    \sigma^{\gamma^{*} {\rm A} \to V {\rm A}}_\textrm{el} \propto 
    \begin{cases}
        A^{4/3}, & \text{outside the saturation region}, \\
        A^{2/3}, & \text{inside the saturation region} .
    \end{cases}
\end{align}
Here transition in or out of the saturation region is obtained by varying $A$ and/or the typical dipole size $r_\perp$. The latter is varied in UPCs by choosing different-size vector mesons, with $J/\psi$ being smaller than $\rho$. 
The inelastic cross section has an $A$-scaling that is dependent on the transverse momentum $p_T$ of the produced quark: 
\begin{align}
    \frac{d\sigma}{d^{2}p_T} \propto 
    \begin{cases}
        A, & p_{T}\gg Q_{s},\\
        A^{2/3} , & p_{T}\ll Q_{s}.
    \end{cases}
\end{align}
The transition into the saturation regime depends on the value of $p_T$ compared to $Q_s$, with the latter depending on $A$ as well. 

Taking the ratio of the elastic vector meson production cross section to the inelastic hadron/jet production cross section, we have the following results. For $J/\psi$ production, assuming that $J/\psi$ is small enough for the process to be outside the saturation region, we first define and calculate the single ratio
\begin{equation}\label{Jpsi R1 classical}
    R_1^{J/\psi} ({\rm A}) \equiv \frac{\sigma^{\gamma^{*} {\rm A} \to J/\psi \, {\rm A}}_\textrm{el} }{\frac{d\sigma}{d^{2}p_T}} \propto
    \begin{cases}
        A^{\frac{1}{3}}, & p_{T}\gg Q_{s},\\
        A^{\frac{2}{3}}, & p_{T}\ll Q_{s}.
    \end{cases}
\end{equation}
If we take the double ratio proposed in \eq{eq3}, the prefactors which depend on $p_T$ only would mostly drop out (given that we evaluate the inelastic cross sections at the same $p_T$) and we obtain the following $A$-scaling for $R_{\rm{UPC}}$,
\begin{equation}\label{eq:Jpsi R2 classical}
    R_{\rm{UPC}}^{J/\psi} \equiv \frac{R_1^{J/\psi} ({\rm A})}{R_1^{J/\psi} (p)}=
    \begin{cases}
        A^{\frac{1}{3}}, & p_{T}\gg Q_{s},\\
        A^{\frac{2}{3}}, & p_{T}\ll Q_{s}.
    \end{cases}
\end{equation}
We see that inside the saturation region for the inclusive cross section ($p_{T}\ll Q_{s}$), the $A$ scaling of $R_{\rm{UPC}}^{J/\psi}$ comes with a higher power of $A$, than outside the saturation region ($p_T \gg Q_s$).

\begin{figure}[ht]
    \centering
    \includegraphics[scale=0.75]{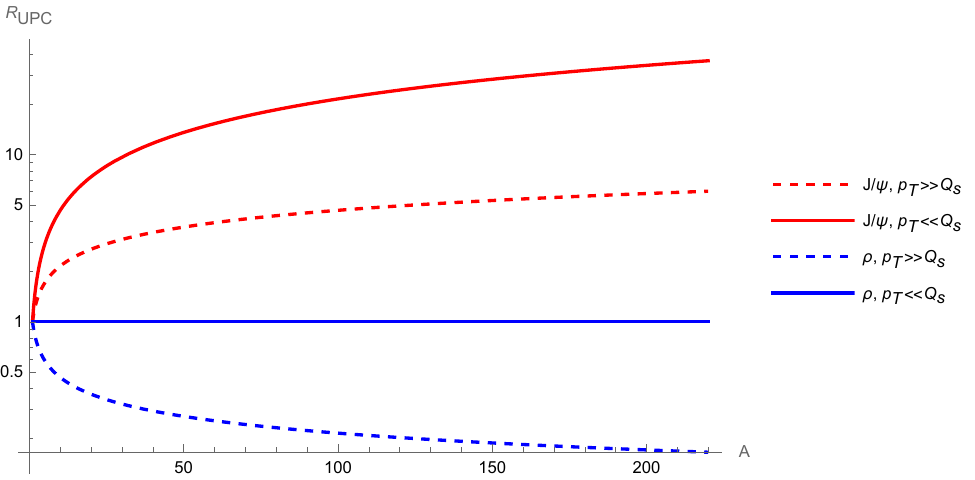}
    \caption{A plot of our approximate estimates for the double ratio $R_{\rm{UPC}}$ as a function of $A$, with $R_{\rm{UPC}}$ given by Eqs.~\eqref{eq:Jpsi R2 classical} and \eqref{eq:rho R2 classical} in the quasi-classical approximation.}
    \label{double ratio vs A classical}
\end{figure}

For the $\rho$ meson production, assuming it is large enough for its elastic production cross section to mainly probe the saturation region, we have the following double ratio due to a different $A$-scaling in the elastic process,
\begin{equation}\label{eq:rho R2 classical}
    R_{\rm{UPC}}^{\rho}=
    \begin{cases}
        A^{-\frac{1}{3}}, & p_{T}\gg Q_{s},\\
        A^0 =1, & p_{T}\ll Q_{s}. 
    \end{cases}
\end{equation}
Again, we see that the power of $A$ is lower outside the saturation region, by the same amount of $1/3$ as in \eq{eq:Jpsi R2 classical}.

Note also, that in going from $J/\psi$ to the larger $\rho$ meson, the powers of $A$ decreased in $R_{\rm{UPC}}$. So the double ratio appears to be sensitive to how one approaches the saturation region: while lowering $p_T$ leads to the double ratio having a higher power of $A$, increasing the dipole (meson) size leads to a decrease of the double ratio for the same $A$ and $p_T$.

A plot of our result for $R_{\rm{UPC}}$ in Eqs.~\eqref{eq:Jpsi R2 classical} and \eqref{eq:rho R2 classical} is shown in \fig{double ratio vs A classical}, demonstrating different $A$ dependence of the double ratio for $J/\psi$ and $\rho$ mesons. Note that a more careful numerical evaluation of the double ratio, akin to that done in Figs.~\ref{Jpsi quasi-classical} and \fig{rho quasi-classical}, is left for the future work: for the realistic inelastic hadron production cross section one would need to augment \eq{eq:inclusive} by the fragmentation functions. In RHIC kinematics, large-$x$ effects may need to be included as well, for the higher $p_T$ values. Additionally, small-$x$ evolution has to be included into the dipole amplitude $N$ for very small $x$ values, as reached in the forward direction at RHIC and LHC. This is what we will do next, to test the validity of our conclusions here.


\subsection{Double ratio with small-$x$ evolution}
\label{sec:evolution}

\subsubsection{Elastic vector meson production}
\label{sec:evolution_elastic}

The results of Section~\ref{sec:classical}, given by Eqs.~\eqref{eq:Jpsi R2 classical} and \eqref{eq:rho R2 classical},
were obtained using the quasi-classical dipole scattering amplitude \eqref{GGM/MV}. However, in practice, radiative corrections need to be incorporated if one wants to obtain a more realistic result, especially at small $x$. In the dipole picture, where the color dipole collides with the target nucleus, the dipole radiates gluons during the process if there is a sufficient scattering energy available. 
These gluon emissions/absorptions result in corrections to the dipole amplitude $N(\mathbf{r},\mathbf{b},Y)$ that are described by the BK/JIMWLK evolution equations \cite{Balitsky:1995ub,Balitsky:1998ya,Kovchegov:1999yj,Kovchegov:1999ua,Jalilian-Marian:1997dw,Jalilian-Marian:1997gr,Weigert:2000gi,Iancu:2001ad,Iancu:2000hn,Ferreiro:2001qy},
making $N(\mathbf{r},\mathbf{b},Y)$ depend on the rapidity interval $Y$ between the dipole and the target nucleus. One consequence of the BK evolution is the extended geometric scaling behavior outside the saturation region (see \cite{Stasto:2000er, Iancu:2002tr, Mueller:2002zm, Munier:2003vc}) and the geometric scaling behavior inside the saturation region (see \cite{Levin:1999mw}). Outside the saturation region 
in the extended geometric region the dipole amplitude takes the form \cite{Munier:2003sj}
\begin{equation}\label{eq:extended geo}
    N(\mathbf{r},\mathbf{b},Y)\propto \left[r_{\perp}^{2}Q_{s}^{2}(\mathbf{b},Y)\right]^{\gamma_{cr}}
\end{equation}
where $\gamma_{cr} \approx 0.6275$ is the solution of
\begin{align}
    \chi (\gamma_{cr}) = \gamma_{cr} \, \frac{d \chi (\gamma_{cr})}{d \gamma_{cr}}
\end{align}
with $\chi(\gamma)$ being the eigenvalue of the LO BFKL kernel given by
\begin{equation}\label{eq:chi}
    \chi(\gamma)=2 \, \psi(1)-\psi\left( \gamma \right)-\psi\left( 1- \gamma \right), \hspace{1cm} \text{where} \hspace{1cm} \psi(\gamma)=\frac{\Gamma'(\gamma)}{\Gamma(\gamma)}.
\end{equation}
With the small-$x$ evolution included, the saturation scale $Q_{s}$ is now a function of rapidity $Y$ given by
\begin{equation}\label{eq:running Qs}
    Q_{s}(\mathbf{b}, Y)= Q_{s} (\mathbf{b}) \, \exp \left[\frac{\alpha_{s}N_{c}}{\pi}\frac{\chi(\gamma_{cr})}{2\gamma_{cr}}Y\right] \approx Q_{s} (\mathbf{b}) \, \exp\left[2.44\frac{\alpha_{s}N_{c}}{\pi}Y\right] 
\end{equation}
where $Q_{s} (\mathbf{b})$ is the quasi--classical saturation scale from \eq{Qs_cl} in Sec.~\ref{sec:classical}. Note that we still have $Q_s (\mathbf{b}, Y)^2 \propto A^{1/3}$.
We see that $N(\mathbf{r},\mathbf{b},Y)$ in \eq{eq:extended geo} depends only on the combined quantity $r_{\perp}Q_{s}(\mathbf{b}, Y)$, instead of depending on two variables $r_\perp$ and $Y$ separately, a behavior known as geometric scaling \cite{Gribov:1981ac, Levin:1999mw, Stasto:2000er, Iancu:2002tr, Mueller:2002zm, Munier:2003vc, Munier:2003sj}. Note that our treatment of the impact parameter $\mathbf{b}$ dependence of the saturation scale is approximate \cite{Kovner:2001bh, Kovner:2002xa, Kovner:2002yt,Golec-Biernat:2003naj,Berger:2010sh,Berger:2011ew}, but should suffice in the spirit of our mainly qualitative arguments here. 

Deep inside the saturation region, the dipole amplitude is given by the Levin-Tuchin (LT) formula \cite{Levin:1999mw}, 
\begin{equation}\label{eq:Levin-Tuchin}
    N(\mathbf{r},\mathbf{b},Y) = 1-S_{0}\exp\left\lbrace-\frac{\gamma_{cr}}{2\, \chi(\gamma_{cr})} \, \ln^2 [r_{\perp}^{2}Q_{s}^{2}(\mathbf{b}, Y)] \right\rbrace ,
\end{equation}
where $S_0$ is a constant specified along the saturation boundary curve $r_{\perp}=1/Q_{s}(\mathbf{b}, Y)$. Again, we see the geometric scaling behavior: $N(\mathbf{r},\mathbf{b},Y)$ depends only on 
the combined quantity $r_{\perp}Q_{s}(\mathbf{b}, Y)$. For a detailed discussion and derivation of these scaling results see Chapter~4.5 in \cite{Kovchegov:2012mbw}.

Let us now examine the $A$-dependence of the elastic vector meson production in UPCs, if we use the dipole amplitude in the scaling region, given by either \eq{eq:extended geo} or \eq{eq:Levin-Tuchin}, in \eq{eq:elastic vmp}. Going back to \eq{eq:elastic vmp}, with the dipole amplitude now depending on $Y$, we apply the same arguments as in Sec.~\ref{sec:classical}. We divide the integration over $r_{\perp}$ and $r_{\perp}'$ into three regions: (\romannumeral 1) $r_{\perp},r_{\perp}' \ll 1/Q_{s}$, (\romannumeral 2) $r_{\perp},r_{\perp}' \gtrsim 1/Q_{s}$, (\romannumeral 3) $r_{\perp} \ll 1/Q_{s},r_{\perp}' \gtrsim 1/Q_{s}$ or vice versa. For $J/\psi$ production, assuming it is small enough to be entirely outside the saturation region, the integral in \eq{eq:elastic vmp} is dominated by region (\romannumeral 1), so we use \eq{eq:extended geo} to approximate the dipole amplitude $N(\mathbf{r},\mathbf{b},Y)$. Since $Q_{s}^{2}(\mathbf{b}, Y) \propto A^{1/3}$, we have $N \sim A^{\gamma_{cr}/3}$, such that $\sigma^{\gamma^{*}{\rm A} \to J/\psi \, {\rm A}}_\textrm{el} \propto A^{\frac{2}{3}(1+\gamma_{cr})}$. For $\rho$ production, assuming the dipole is large enough to be inside the saturation region, the integral in \eq{eq:elastic vmp} is dominated by region (\romannumeral 2), so we use the approximation that $N(\mathbf{r},\mathbf{b},Y)\approx 1$ (instead of the LT formula) for convenience. We would then have $\sigma^{\gamma^{*}{\rm A}\to \rho \, {\rm A}}_\textrm{el} \propto A^{\frac{2}{3}}$, same as in the quasi--classical regime of Sec.~\ref{sec:classical}.

\begin{figure}[ht]
    \centering
    \includegraphics[scale=0.88]{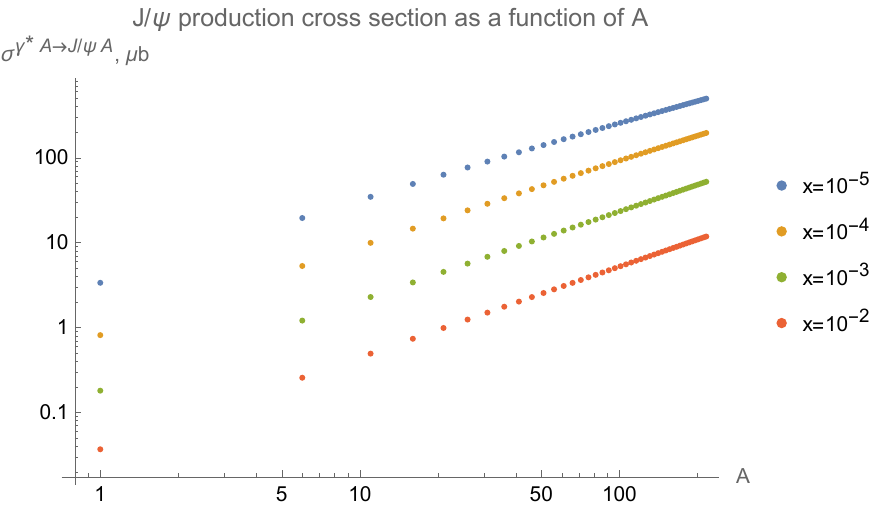}
    \caption{A log-log plot of the elastic $J/\psi$ production cross section versus the atomic number $A$ of the target nucleus for four different $x$-values, $x = \{ 10^{-2}, 10^{-3}, 10^{-4}, 10^{-5} \}$. The dipole amplitude used here includes small-$x$ evolution and is given in \eq{eq: num dip amp evo}.}
    \label{fig:Jpsi evolution}
\end{figure}

Since the $r_{\perp},r_{\perp}'$ integrals in \eq{eq:elastic vmp} run over the entire range of $r_{\perp},r_{\perp}' \in [0,\infty]$, we would like to perform a numerical integration to obtain a more precise estimate of the $A$-scaling for the elastic cross sections. This time, we use a dipole amplitude that is a solution of the BK evolution equation. Unfortunately, the exact analytic solution to the BK equation is not known, so we have to either solve the BK equation numerically, or resort to an approximation. In our present exploratory study, we chose the latter, and approximate the dipole scattering amplitude 
by piecing together the extended geometric scaling behavior formula \eqref{eq:extended geo} outside the saturation region ($r_{\perp}\lesssim 1/Q_{s}$) with the LT formula \eqref{eq:Levin-Tuchin} inside the saturated regime ($r_{\perp}\gtrsim 1/Q_{s}$). In addition, we require the dipole amplitude to be continuous at $r_{\perp}=1/Q_{s}$. We thus write 
\begin{align}\label{eq: num dip amp evo}
    N(\mathbf{r},\mathbf{b},Y) = & \ (1-S_{0}) \,[r_{\perp}^{2}Q_{s}^{2}(\mathbf{b},Y)]^{\gamma_{cr}} \, \theta \left(1 - r_{\perp} Q_{s} (\mathbf{b},Y) \right) \\ 
    & + \left[ 1-S_{0}\exp\left\lbrace-\frac{\gamma_{cr}}{2\, \chi(\gamma_{cr})} \, \ln^2 [r_{\perp}^{2}Q_{s}^{2}(\mathbf{b},Y)] \right\rbrace \right] \, \theta \left( r_{\perp} Q_{s} (\mathbf{b}, Y) - 1 \right). \notag 
\end{align}
Here $S_{0}$ is some constant determined by the initial condition for the forward scattering amplitude. Since the curve $Q_{s}(Y)$ which describes the rapidity dependence of the saturation scale $Q_{s}$ is determined by requiring the dipole amplitude $N(1/Q_{s}(Y),Y)$ to be constant along this curve, we have the freedom to choose what this constant value is as long as it is of the order of 1. This, in turn, leads to the freedom in $S_{0}$. A typical choice of $S_{0}$ motivated by the GGM/MV model (cf. \eq{GGM/MV}) is to take $S_{0}=e^{-\frac{1}{4}}.$ Using this $S_{0}$ in \eq{eq: num dip amp evo} and employing the resulting dipole amplitude to perform the numerical integrations in \eq{eq:elastic vmp} leads to the log-log plot of $J/\psi$ elastic production cross section versus $A$ depicted in \fig{fig:Jpsi evolution} for a sample of small $x$ values, $x = \{ 10^{-2}, 10^{-3}, 10^{-4}, 10^{-5} \}$. The saturation scale is given by \eq{eq:running Qs} with $Q_s (\mathbf{b})$ from \eq{Qs_cl}. We put $\as = 0.12$ in \eq{eq:running Qs} to account for the effects of higher-order corrections on the $x$-dependence of the saturation scale in our leading-order formalism. (Note that here $Y = \ln (x_0/x)$ with $x_0 = 2 \times 10^{-3}$.) 

In \fig{fig:Jpsi evolution} we see the expected trend of the scattering cross section increasing with increasing scattering energy (with decreasing $x$). The slopes $\alpha$ from \eq{alpha}, calculated again using pairs of adjacent data points in \fig{fig:Jpsi evolution}, is summarized in Table~\ref{Jpsi slope}.  We observe that the slope at larger values of $x$,  $x=10^{-2}$ or $x=10^{-3}$, is very close to the estimated value of $\frac{2}{3}(1+\gamma_{cr}) \approx 1.085$ in region (i) considered above (outside of the saturation region), indicating that the elastic $J/\psi$ production process for larger $x$ is indeed non-saturated in our picture, but is still inside the geometric scaling region. However, as we go to lower $x$, the scaling power $\alpha$ decreases significantly, becoming much smaller than in the quasi-classical case, while still remaining significantly above the fully-saturated value of $2/3$. 
This implies that the increase of $Q_{s}$ at small $x$ moves the $J/\psi$ production process more into the saturated region.

\begin{table}[h]
    \centering
    \begin{tabular}{|c|c|c|}
        \hline
        $x$ & slope at $A=1$ & slope at $A=208$\\
        \hline
        $10^{-2}$ & 1.08 & 1.06\\
        \hline
        $10^{-3}$ & 1.06 & 1.04\\
        \hline
        $10^{-4}$ & 1.05 & 0.96\\
        \hline
        $10^{-5}$ & 0.98 & 0.85\\
        \hline
    \end{tabular}
    \caption{The effective scaling power $\alpha$ of $A$ for the $J/\psi$ elastic production cross section (the slopes in \fig{fig:Jpsi evolution}) for different $x$ values: $\sigma_\textrm{el}^{\gamma^{*} {\rm A} \to J/\psi \, {\rm A}}\propto A^\alpha$.}
    \label{Jpsi slope}
\end{table}

\begin{figure}[h]
    \centering
    \includegraphics[scale=0.8]{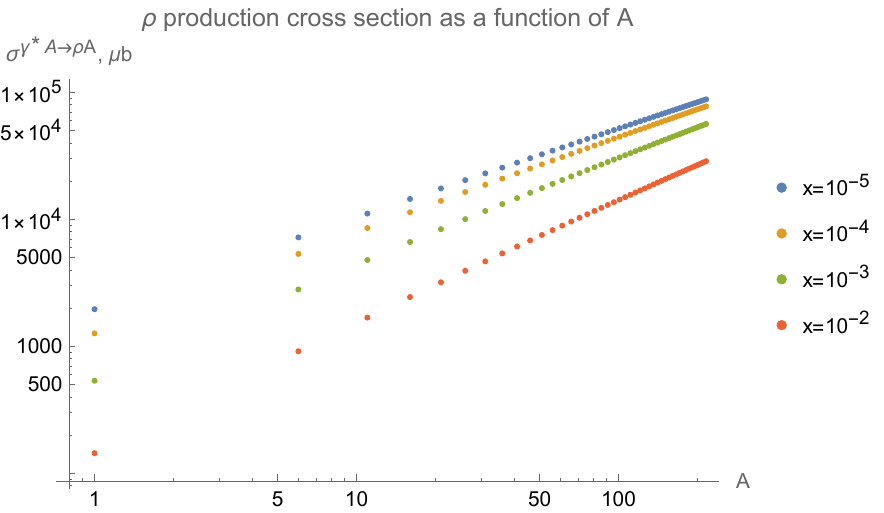}
    \caption{A log-log plot of the elastic $\rho$ meson production cross section as a function of the atomic number $A$ of the target nucleus for four different $x$-values, $x = \{ 10^{-2}, 10^{-3}, 10^{-4}, 10^{-5} \}$. The dipole amplitude used here includes small-$x$ evolution and is given in \eq{eq: num dip amp evo}.}
    \label{fig:rho evolution}
\end{figure}

Performing a similar numerical integration by using the amplitude \eqref{eq: num dip amp evo} in \eq{eq:elastic vmp}, but now for the elastic $\rho$ meson production yields the log-log plot shown in \fig{fig:rho evolution}. The slope of the data points is summarized in Table \ref{rho slope}. The scaling powers $\alpha$ are
smaller than those for $J/\psi$ production, which is expected since saturation effects are stronger for the larger mesons like $\rho$. For small $x$ values such as $x=10^{-5}$, the slope is very close to the $2/3$ scaling power for a fully saturated process (region (ii) above). However for larger $x$, even for $x=10^{-3}$, it deviates quite noticeably from the above estimate of $\alpha=2/3$ for a perfectly saturated process, and $\alpha$ is larger than that extracted from \fig{rho quasi-classical} for the quasi--classical case.   
This is because $N(\mathbf{r},\mathbf{b},Y)$ approaches 1 more slowly in the LT formula (cf. \eq{eq:Levin-Tuchin}) than in the quasi-classical case (cf. \eq{eq: num dip amp classical}): in the latter case, $N$ approaches 1 exponentially in $r_\perp^2$ instead of the exponential of $\ln^2 [r_\perp^2 \, Q_s^2]$ in the former case (the LT formula). Hence, the saturation effects in terms of $A$-scaling manifest themselves less when evolution corrections are included.

\begin{table}[h]
    \centering
    \begin{tabular}{|c|c|c|}
    \hline
        $x$ & slope at $A=1$ & slope at $A=208$\\
        \hline
        $10^{-2}$ & 1.03 & 0.90\\
        \hline
        $10^{-3}$ & 0.92 & 0.79\\
        \hline
        $10^{-4}$ & 0.80 & 0.72\\
        \hline
        $10^{-5}$ & 0.73 & 0.69\\
        \hline
    \end{tabular}
    \caption{The effective scaling power $\alpha$ of $A$ for the $\rho$ meson elastic production cross section (the slopes in \fig{fig:rho evolution}) for different $x$ values: $\sigma_\textrm{el}^{\gamma^{*} {\rm A} \to \rho \, {\rm A}}\propto A^\alpha$.
    }
    \label{rho slope}
\end{table}

Still, the overall pattern for the elastic vector meson production remains the same after inclusion of small-$x$ evolution corrections: the effect of saturation is to reduce the power of $A$ in the cross section. Again, this reduction can be achieved by either increasing $A$ or by increasing the dipole size (in going from $J/\psi$ to $\rho$). Additionally, with small-$x$ evolution included, the reduction of the power of $A$ can be achieved by going to smaller $x$.


\subsubsection{Inclusive production cross section}
\label{sec:evolution_incl}

Similar to what we did for the elastic vector meson production, we can include small-$x$ evolution in the inclusive production cross section \eqref{eq:inclusive} by using the dipole amplitude $N(\mathbf{r},\mathbf{b},Y)$ 
given by the solution of the BK equation. Again, we separately consider the $p_{T}\gg Q_{s}$ and $\Lambda_{QCD}\ll p_{T}\ll Q_{s}$ cases.

Outside the saturation region, when $p_{T}\gg Q_{s}$, the contribution to the $r_{\perp}$-integral in \eq{eq:inclusive} mainly comes from the region where $r_{\perp}\ll 1/p_{T}\ll 1/Q_{s}$, so we are in the extended geometric scaling region. Hence we approximate $N(\mathbf{r},\mathbf{b},Y)$ by \eq{eq:extended geo}, which leads to $N \propto A^{\frac{1}{3}\gamma_{cr}}$, and consequently 
\begin{align}\label{incl_out_evol}
    \frac{d\sigma}{d^{2}p_T} \bigg|_{p_T \gg Q_s} \propto A^{\frac{2}{3}+\frac{1}{3}\gamma_{cr}} .
\end{align}

Inside the saturation region, when $\Lambda_{QCD}\ll p_{T}\ll Q_{s}$, the contribution to the integral in \eq{eq:inclusive} mainly comes from the region where $1/Q_{s}\lesssim r_{\perp}\lesssim 1/p_{T}$, so we are probing the saturation region. Hence we approximate $N(\mathbf{r},\mathbf{b},Y)\approx1$, and consequently 
\begin{align}\label{incl_in_eval}
    \frac{d\sigma}{d^{2}p_T} \bigg|_{p_T \lesssim Q_s} \propto A^{2/3} ,
\end{align} 
same as in the quasi--classical case.


\subsubsection{$A$-scaling for the double ratio for $J/\psi$ and $\rho$ production}

Let us combine the results above for the elastic vector meson production and the inclusive hadron/jet production. The analytic results of Sec.~\ref{sec:evolution_elastic} can be summarized as
\begin{align}
    \sigma^{\gamma^{*} {\rm A} \to V{\rm A}}_\textrm{el} \propto 
    \begin{cases}
        A^{\frac{2}{3} (1+\gamma_{cr})}, & \text{outside the saturation region}, \\
        A^{\frac{2}{3}}, & \text{inside the saturation region} .
    \end{cases}
\end{align}
Similarly, the results of \sec{sec:evolution_incl} are
\begin{align}
    \frac{d\sigma}{d^{2}p_T} \propto 
    \begin{cases}
        A^{\frac{2}{3}+\frac{1}{3}\gamma_{cr}}, & p_{T}\gg Q_{s},\\
        A^{\frac{2}{3}} , & p_{T}\ll Q_{s}.
    \end{cases}
\end{align}

We see that when the small-$x$ evolution is taken into account, we have the following single ratio of the elastic $J/\psi$ production cross section to the inelastic jet/hadron production cross section:
\begin{equation}\label{Jpsi R1 evolution}
    R_1^{J/\psi} ({\rm A})  = \frac{\sigma^{\gamma^{*} {\rm A} \to J/\psi \, {\rm A}}_\textrm{el} }{\frac{d\sigma}{d^2 p_T}} \propto
    \begin{cases}
        A^{\frac{1}{3}\gamma_{cr}}, & p_{T}\gg Q_{s},\\
        A^{\frac{2}{3}\gamma_{cr}}, & p_{T}\ll Q_{s}.
    \end{cases}
\end{equation}
Taking the double ratio gives
\begin{equation}\label{eq:Jpsi R2 evolution}
    R_{\rm{UPC}}^{J/\psi}=
    \begin{cases}
        A^{\frac{1}{3}\gamma_{cr}}, & p_{T}\gg Q_{s},\\
        A^{\frac{2}{3}\gamma_{cr}}, & p_{T}\ll Q_{s}.
    \end{cases}
\end{equation}
For the $\rho$ meson, 
the double ratio now is
\begin{equation}\label{eq:rho R2 evolution}
    R_{\rm{UPC}}^{\rho}=
    \begin{cases}
        A^{-\frac{1}{3}\gamma_{cr}}, & p_{T}\gg Q_{s},\\
        A^0 = 1, & p_{T}\ll Q_{s}.
    \end{cases}
\end{equation}

\begin{figure}[h]
    \centering
    \includegraphics[scale=0.75]{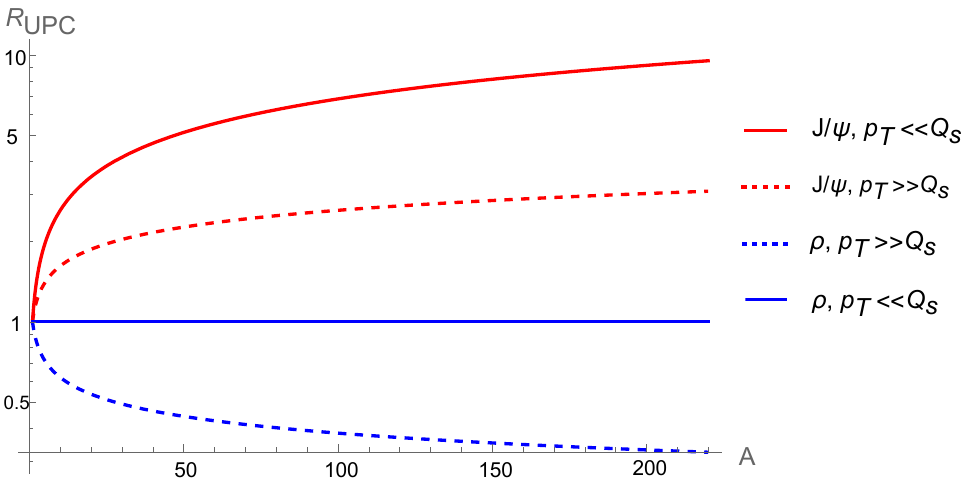}
    \caption{The double ratio $R_{\rm{UPC}}$ plotted versus $A$ for the case where we include LO small-$x$ evolution into the elastic and inclusive cross sections. The figure illustrates the results in Eqs.~\eqref{eq:Jpsi R2 evolution} and \eqref{eq:rho R2 evolution}.}
    \label{double ratio vs A evolution}
\end{figure}

Using $\gamma_{cr}=0.6275$, we plot our results from Eqs.~\eqref{eq:Jpsi R2 evolution} and \eqref{eq:rho R2 evolution} for $R_{\rm{UPC}}$ versus $A$ in \fig{double ratio vs A evolution}. Figure~\ref{double ratio vs A evolution} is qualitatively similar to \fig{double ratio vs A classical}, with the somewhat lower powers of $A$ for some of the curves in \fig{double ratio vs A evolution}. Further phenomenological analysis of the double ratio, which may  involve solving the evolution equations at next-to-leading (NLO) order \cite{Balitsky:2008zza, Kovner:2013ona, Grabovsky:2013mba, Balitsky:2014mca}, along with constructing NLO expressions for the corresponding elastic and inclusive cross sections (see, e.g., \cite{Boussarie:2016bkq}), is left for the future work improving on our present exploratory study.

%
\section{Future opportunities at RHIC and LHC}
%
\label{sec:future}

In the near term at RHIC, there are two running experiments, sPHENIX and STAR, taking high-luminosity Au$+$Au and p$+$p data at 200 GeV center-of-mass energy for Run 2024 and Run 2025. There is a possibility 
of a $p+$Au run in 2024, if condition allows, which will be valuable for the baseline measurement of scattering on the proton. (Note that STAR performed a $d+$Au UPC $J/\psi$ measurement~\cite{STAR:2021wwq} but the data do not have a dedicated jet trigger.) These two runs will provide the highest-statistics UPC samples for the entire RHIC era. 

Specifically, the sPHENIX experiment is a brand new detector with the capability of performing UPC measurements, while the STAR experiment has a well-established UPC program and the newly installed forward detector system will provide a large extension in the kinematic phase space for UPCs. The advantage of UPC measurements at RHIC is the unique kinematics that transitions from high-$x$ to low-$x$, where saturation physics is expected to start playing a role ($x<0.01$). Furthermore, with the extension of the STAR forward rapidity coverage, measurement of the UPC jet or semi-inclusive hadron photoproduction events may become possible with dedicated triggers. Together with VM production, e.g., $\rho$ and $J/\psi$, the single and double ratio observables we proposed here can be measured experimentally. Due to the excellent tracking system in STAR, it is possible to measure the photoproduction of $\phi$ meson with soft kaon decays for the first time (average momentum for the kaon daughter is around 100 $\rm MeV/c$), which will provide an extra handle on the mass or dipole size dependence of our proposed observables. 

At the LHC, the Run 3 has just started with three Heavy-Ion runs scheduled. All four LHC experiments have been participating in UPC measurements~\cite{Khachatryan:2016qhq,Abelev:2012ba,ALICE:2020ugp,ALICE:2021jnv,ALICE:2021tyx,ALICE:2021gpt,LHCb:2021hoq,LHCb:2022ahs,CMS:2023snh,ALICE:2023jgu}. Due to the higher center-of-mass energy, the kinematic phase space for most experiments is expected to be deep into the saturation regime. There is even a hint of saturation phenomenon recently reported by the CMS experiment~\cite{CMS:2022nnw}, although the interpretation is not entirely clear. The ATLAS experiment, although it has not yet reported any VM photoproduction measurement, has pioneered UPC jets study in the past years based on the data from Run 2. There is a good experimental opportunity for the ATLAS experiment to simultaneously measure elastic VM and inclusive jet productions in UPCs to study the single and double ratios proposed in this paper. In addition, an Oxygen-Oxygen collision run is possible at the LHC in Run 3, which would provide an extra point in the $A$-dependent ($A=16$) UPC measurement. In the longer term, when the LHC Run-4 data will become available, there will be still a few years before the operation of the EIC. At Run-4, the ALICE experiment may have a Forward Calorimeter (FoCal) upgrade \cite{ALICE:2023fov} that will be able to probe $x$ down to $x \sim 10^{-6}$. This will be another great opportunity for testing our proposed measurements quantitatively. 

%
\section{Summary and Outlook}
%
\label{sec:conclusions}

In this paper, inspired by the double ratio of the diffractive to total cross section in DIS to be measured at the EIC~\cite{Accardi:2012qut}, we have proposed a novel double ratio, $R_{\rm{UPC}} ({\rm A})$ in \eq{eq3}, that is also sensitive to saturation effects and can be measured in the UPCs at RHIC and LHC. More precisely, we have performed a calculation of $R_{\rm{UPC}} ({\rm A})$ for $J/\psi$ and $\rho$ meson production in the saturation framework. Within this framework, we have considered the quasi--classical approximation first, and included the small-$x$ evolution later. Our results for the double ratios are given in Eqs.~\eqref{eq:Jpsi R2 classical} and \eqref{eq:rho R2 classical} in the quasi--classical case and by Eqs.~\eqref{eq:Jpsi R2 evolution} and \eqref{eq:rho R2 evolution} for the small-$x$ evolution case. In both calculations, we observe a different dependence of $R_{\rm{UPC}} ({\rm A})$ on the atomic number $A$ inside and outside the saturation region. 

Whether we are probing the physics inside or outside the saturation region is determined by the transverse momentum $p_T$ of the hadron or jet: for $p_T \lesssim Q_s$ the inclusive production cross section in the denominator of the single ratio $R_1$ (defined in \eq{Jpsi R1 classical}) is largely probing the physics inside the saturation region, while for $p_T \gg Q_s$ it is probing the physics outside the saturation region. The size of the produced meson in the elastic cross section also affects whether saturation region is probed or not in the elastic cross section in the numerator of $R_1$. In the case of a relatively compact meson production,  for $J/\psi$, the double ratio $R_{\rm{UPC}} ({\rm A})$ grows faster with $A$ inside the saturation region ($p_T \lesssim Q_s$) than outside ($p_T > Q_s$), as follows from Eqs.~\eqref{eq:Jpsi R2 classical} and \eqref{eq:Jpsi R2 evolution}.  For a larger meson, like $\rho$, the double ratio $R_{\rm{UPC}} ({\rm A})$ decreases with $A$ outside the saturation region ($p_T > Q_s$), and remains relatively flat in $A$ inside the saturation region ($p_T \lesssim Q_s$), as per Eqs.~\eqref{eq:rho R2 classical} and  \eqref{eq:rho R2 evolution}. 

If our, admittedly preliminary, estimates are confirmed by the future more detailed calculations, possibly involving NLO corrections to the cross sections and evolution equations, then one may be able to conclude that the $A$-dependence of $R_{\rm{UPC}} ({\rm A})$ would be a novel signal of saturation physics, to be explored in the current and future UPC measurements at RHIC and LHC.

To prepare for the future experiments and to make sure the behavior predicted here is unique to saturation physics, one has to compare the saturation predictions for $R_{\rm UPC}$ to those coming from other calculations. In particular, similar to~\cite{Accardi:2012qut}, one may consider the LTS model \cite{Kopeliovich:1999am, Kopeliovich:2002yv, Frankfurt:2003gx, Frankfurt:2011cs}. While our $R_{\rm{UPC}} ({\rm A})$ has not been calculated in the LTS model yet, a similar (single) ratio of the diffractive to inclusive di-jet production cross sections in UPCs was studied recently in \cite{Guzey:2020ehb}. The conclusion of \cite{Guzey:2020ehb} appears to indicate that the diffractive to inclusive di-jet ratio in the LTS picture decreases with the atomic number $A$ when comparing ultra-peripheral $p+p$ and A$+$A collisions. This means that the corresponding double ratio, if constructed, would be smaller than one and decreasing with $A$. This appears to be different from the $A$-dependence we obtained above for $R_{\rm UPC}$ in most (though not all) regimes considered, making us cautiously optimistic that an LTS-based calculation for $R_{\rm UPC}$ would lead to a significantly different $A$-dependence from that in the saturation picture, potentially allowing one to distinguish between the two approaches using the future UPCs data.


\section*{Acknowledgments}

\label{sec:acknowledgement}

The authors would like to thank Peter Steinberg for information about the UPC program in the ATLAS experiment. We would like to thank the Brookhaven EIC group for general discussion on UPCs and saturation physics. 

This material is based upon work supported by the U.S. Department of Energy, Office of Science, Office of Nuclear Physics under Award Number DE-SC0004286 and the work performed within the framework of the Saturated Glue (SURGE) Topical Theory Collaboration  (YK and HS). 

The work of ZT is supported by the U.S. Department of Energy under Award DE-SC0012704 and the Laboratory Directed Research LDRD-23-050 project.



 \appendix
 \section{Calculational details}
 
 \label{A}

Using the following integral formulae,
\begin{subequations}
\begin{align}
    & \int\frac{d^{2}q_{\perp}}{q_{\perp}^{2} + a_{f}^{2}}e^{i\mathbf{q}\cdot\mathbf{x}}=2\pi \, K_{0}(a_{f}x_{\perp}),\label{eq35}\\ 
    & \int d^{2}q_{\perp}\frac{\mathbf{q}}{q_{\perp}^{2} + a_{f}^{2}}e^{i\mathbf{q}\cdot\mathbf{x}}=2\pi \, i \, a_{f} \, K_{1}(a_{f}x_{\perp}) \, \frac{\mathbf{x}}{x_{\perp}},\label{eq36}
\end{align}
\end{subequations}
we can perform some of the integrals in \eq{eq34} term by term:
\begin{subequations}\label{integrals}
\begin{align}
    &\int d^{2}x_{\perp}^{\prime}d^{2}y_{\perp}e^{i\mathbf{p}\cdot(\mathbf{x}^{\prime}-\mathbf{y})}\int d^{2}x_{\perp}e^{-i\mathbf{p}\cdot(\mathbf{x}-\mathbf{y})}\left\lbrace[z^{2}+(1-z)^{2}]a_{f}^{2}K_{1}(|\mathbf{x}^{\prime}-\mathbf{y}|a_{f})\frac{\mathbf{x}^{\prime}-\mathbf{y}}{|\mathbf{x}^{\prime}-\mathbf{y}|}\cdot\frac{1}{2\pi ia_{f}}\right. \\ 
    &\ \ \ \ \ \ \ \ \ \ \ \ \ \ \ \times\left.\int d^{2}q_{\perp}\frac{\mathbf{q}}{q_{\perp}^{2}+a_{f}^{2}}e^{i\mathbf{q}\cdot(\mathbf{x}-\mathbf{y})}+m_{f}^{2}K_{0}(|\mathbf{x}^{\prime}-\mathbf{y}|a_{f})\frac{1}{2\pi}\int\frac{d^{2}q_{\perp}}{q_{\perp}^{2}+a_{f}^{2}}e^{i\mathbf{q}\cdot(\mathbf{x}-\mathbf{y})}\right\rbrace  N \left( \mathbf{x}'-\mathbf{y} , \frac{\mathbf{x}'+\mathbf{y}}{2} , Y \right)  \nonumber\\=
    &\int d^{2}x_{\perp}^{\prime}d^{2}y_{\perp}e^{i\mathbf{p}\cdot(\mathbf{x}^{\prime}-\mathbf{y})}\left\lbrace[z^{2}+(1-z)^{2}]a_{f}K_{1}(|\mathbf{x}^{\prime}-\mathbf{y}|a_{f})(-2\pi i)\frac{\mathbf{p}\cdot(\mathbf{x}^{\prime}-\mathbf{y})}{(p_{\perp}^{2}+a_{f}^{2})|\mathbf{x}^{\prime}-\mathbf{y}|}\right.\nonumber\\ 
    &\ \ \ \ \ \ \ \ \ \ \ \ \ \ \ \ \ \ \ \ \ \ \ \ \ \ \ \ \ \ \ \ \ \ \ \ \ \ \ \ \ \ \ \ \ \ \ \ \ \ \ \ \ \ \ \ \ \ \ \ \ \ \ +\left.\frac{2\pi m_{f}^{2}}{p_{\perp}^{2}+a_{f}^{2}}K_{0}(|\mathbf{x}^{\prime}-\mathbf{y}|a_{f})\right\rbrace  N \left( \mathbf{x}'-\mathbf{y} , \frac{\mathbf{x}'+\mathbf{y}}{2} , Y \right)  \nonumber \\ 
    & = 2\pi\int d^{2}x_{\perp}d^{2}x_{\perp}^{\prime}e^{-i\mathbf{p}\cdot(\mathbf{x}-\mathbf{x}^{\prime})}\left\lbrace[z^{2}+(1-z)^{2}] \, a_{f} \, K_{1}(|\mathbf{x}-\mathbf{x}^{\prime}|a_{f}) \, i \, \frac{\mathbf{p}\cdot(\mathbf{x}-\mathbf{x}^{\prime})}{(p_{\perp}^{2}+a_{f}^{2})|\mathbf{x}-\mathbf{x}^{\prime}|}+\frac{m_{f}^{2}}{p_{\perp}^{2}+a_{f}^{2}}K_{0}(|\mathbf{x}-\mathbf{x}^{\prime}|a_{f})\right\rbrace \notag \\ 
    & \hspace*{5cm} \times \, N \left( \mathbf{x}'-\mathbf{x} , \frac{\mathbf{x}+\mathbf{x}'}{2} , Y \right), \notag \\
    &\int d^{2}x_{\perp}d^{2}y_{\perp}e^{-i\mathbf{p}\cdot(\mathbf{x}-\mathbf{y})}\int d^{2}x_{\perp}^{\prime}e^{i\mathbf{p}\cdot(\mathbf{x}^{\prime}-\mathbf{y})}\left\lbrace\left[z^{2}+(1-z)^{2}\right]a_{f}^{2}K_{1}(|\mathbf{x}-\mathbf{y}|a_{f})\frac{\mathbf{x}-\mathbf{y}}{|\mathbf{x}-\mathbf{y}|}\cdot\frac{1}{2\pi ia_{f}}\right. \\ &\hspace{3cm}\times\left.\int d^{2}q_{\perp}\frac{\mathbf{q}}{q_{\perp}^{2}+a_{f}^{2}}e^{i\mathbf{q}\cdot(\mathbf{x}^{\prime}-\mathbf{y})}+m_{f}^{2}K_{0}(|\mathbf{x}-\mathbf{y}|a_{f})\frac{1}{2\pi}\int\frac{d^{2}q_{\perp}}{q_{\perp}^{2}+a_{f}^{2}}e^{i\mathbf{q}\cdot(\mathbf{x}^{\prime}-\mathbf{y})}\right\rbrace N \left( \mathbf{x}-\mathbf{y} , \frac{\mathbf{x}+\mathbf{y}}{2} , Y \right) \nonumber\\ 
    & = \int d^{2}x_{\perp}d^{2}y_{\perp}e^{-i\mathbf{p}\cdot(\mathbf{x}-\mathbf{y})}\left\lbrace[z^{2}+(1-z)^{2}]a_{f}K_{1}(|\mathbf{x}-\mathbf{y}|a_{f})(2\pi i)\frac{\mathbf{p}\cdot(\mathbf{x}-\mathbf{y})}{(p_{\perp}^{2}+a_{f}^{2})|\mathbf{x}-\mathbf{y}|}+\frac{2\pi m_{f}^{2}}{p_{\perp}^{2}+a_{f}^{2}}K_{0}(|\mathbf{x}-\mathbf{y}|a_{f})\right\rbrace \notag \\ 
    & \hspace*{5cm} \times \, N \left( \mathbf{x}-\mathbf{y} , \frac{\mathbf{x}+\mathbf{y}}{2} , Y \right) \nonumber\\ 
    & = 2\pi\int d^{2}x_{\prime}d^{2}x_{\perp}^{\prime}e^{-i\mathbf{p}\cdot(\mathbf{x}-\mathbf{x}^{\prime})}\left\lbrace[z^{2}+(1-z)^{2}]a_{f}K_{1}(|\mathbf{x}-\mathbf{x}^{\prime}|a_{f})i\frac{\mathbf{p}\cdot(\mathbf{x}-\mathbf{x}^{\prime})}{(p_{\perp}^{2}+a_{f}^{2})|\mathbf{x}-\mathbf{x}^{\prime}|}+\frac{m_{f}^{2}}{p_{\perp}^{2}+a_{f}^{2}}K_{0}(|\mathbf{x}-\mathbf{x}^{\prime}|a_{f})\right\rbrace \notag \\ 
    & \hspace*{5cm} \times \, N \left( \mathbf{x}-\mathbf{x}' , \frac{\mathbf{x}+\mathbf{x}'}{2} , Y \right) \nonumber , \\
    &\int d^{2}x_{\perp}d^{2}x_{\perp}^{\prime}e^{-i\mathbf{p}\cdot(\mathbf{x}-\mathbf{x}^{\prime})}\int d^{2}y_{\perp}\left\lbrace[z^{2}+(1-z)^{2}]a_{f}^{2}\left(\frac{1}{2\pi ia_{f}}\right)^{2}\int d^{2}q_{\perp}\frac{\mathbf{q}}{q_{\perp}^{2}+a_{f}^{2}}e^{i\mathbf{q\cdot(\mathbf{x}-\mathbf{y})}} \right. \\ 
    & \times \, \int d^{2}q_{\perp}^{\prime}\frac{\mathbf{q}^{\prime}}{q_{\perp}^{\prime 2}+a_{f}^{2}}e^{i\mathbf{q}^{\prime}\cdot(\mathbf{x}^{\prime}-\mathbf{y})} \left.+m_{f}^{2}\left(\frac{1}{2\pi}\right)^{2}\int\frac{d^{2}q_{\perp}}{q_{\perp}^{2}+a_{f}^{2}}e^{i\mathbf{q}\cdot(\mathbf{x}-\mathbf{y})}\int\frac{d^{2}q_{\perp}^{\prime}}{q_{\perp}^{\prime 2}+a_{f}^{2}}e^{i\mathbf{q}^{\prime}\cdot(\mathbf{x}^{\prime}-\mathbf{y})}\right\rbrace N \left( \mathbf{x}-\mathbf{x}' , \frac{\mathbf{x}+\mathbf{x}'}{2} , Y \right) \nonumber\\ 
    & = \int d^{2}x_{\perp}d^{2}x_{\perp}^{\prime}e^{-i\mathbf{p}\cdot(\mathbf{x}-\mathbf{x}^{\prime})}\left\lbrace[z^{2}+(1-z)^{2}]\int d^{2}q_{\perp}\frac{q_{\perp}^{2}}{(q_{\perp}^{2}+a_{f}^{2})^{2}}e^{i\mathbf{q}\cdot(\mathbf{x}-\mathbf{x}^{\prime})}+m_{f}^{2}\int\frac{d^{2}q_{\perp}}{(q_{\perp}^{2}+a_{f}^{2})^{2}}e^{i\mathbf{q}\cdot(\mathbf{x}-\mathbf{x}^{\prime})}\right\rbrace \notag \\ 
    & \hspace{5cm}  \times \, N \left( \mathbf{x}-\mathbf{x}' , \frac{\mathbf{x}+\mathbf{x}'}{2} , Y \right) \nonumber\\ 
    & = \int d^{2}x_{\perp}d^{2}x_{\perp}^{\prime}e^{-i\mathbf{p}\cdot(\mathbf{x}-\mathbf{x}^{\prime})}\left\lbrace[z^{2}+(1-z)^{2}]\pi\Big[ 2 \, K_{0}(|\mathbf{x}-\mathbf{x}^{\prime}|a_{f})-|\mathbf{x}-\mathbf{x}^{\prime}|a_{f} \, K_{1}(|\mathbf{x}-\mathbf{x}^{\prime}|a_{f})\Big] \right. \notag \\ 
    & \hspace{5cm}  \left. +m_{f}^{2}\frac{\pi|\mathbf{x}-\mathbf{x}^{\prime}|}{a_{f}}K_{1}(|\mathbf{x}-\mathbf{x}^{\prime}|a_{f})\right\rbrace \, N \left( \mathbf{x}-\mathbf{x}' , \frac{\mathbf{x}+\mathbf{x}'}{2} , Y \right) . \nonumber
\end{align}
\end{subequations}
Substituting Eqs.~\eqref{integrals} into \eq{eq34}, relabeling variables appropriately, and employing \eq{tr_symm}, leads to \eq{eq:inclusive}. 


\bibliographystyle{JHEP}
\bibliography{references}

\end{document}